\documentclass[twocolumn]{aastex62}

\usepackage{tikz}
\usepackage{CJK}
\usepackage{multirow}
\usepackage{amsmath,amssymb,amstext}
\usepackage{graphicx}
\usepackage{ulem}
\newcommand{\Lzo}{\ensuremath{ L_{z0}}}
\newcommand{\Lz}{\ensuremath{ L_{z}}}
\newcommand{\ppm}{\ensuremath{ \mathbf{p_m}  }}
\newcommand{\fe}{\ensuremath{\mathrm{[Fe/H]}}}
\newcommand{\dfe}{\ensuremath{\nabla\mathrm{[Fe/H]}}}	% metallicity grad at sun

\newcommand{\Ro}{\ensuremath{R_0}}

\newcommand{\tsfr}{\ensuremath{\tau_{\tiny{SFR}}}}
\newcommand{\srm}{\ensuremath{\sigma_\mathrm{Lz}}}

\newcommand{\gfe}{\ensuremath{{\gamma_\mathrm{[Fe/H]}}}}
	
\newcommand{\vcirc}{\ensuremath{v_\mathrm{circ}}}
\newcommand{\rdo}{\ensuremath{R_{d0}}}
\newcommand{\xio}{\ensuremath{ x_{io}}}

\newcommand{\update}[1]{{\textcolor{black}{#1}}}

\definecolor{mygreen}{HTML}{006d2c}

\shorttitle{Much Migration, Little Heating in the MW Disk}
\shortauthors{Frankel et al.}

\begin{document} 
\begin{CJK*}{UTF8}{gbsn}

\author[0000-0002-6411-8695	]{Neige Frankel}\email{frankel@mpia.de}

\affiliation{Max Planck Institute for Astronomy, K\"onigstuhl 17, D-69117 Heidelberg, Germany}

\author[0000-0003-4593-6788]{Jason Sanders}
\affiliation{Institute of Astronomy, University of Cambridge, Madingley Road, Cambridge CB3 0HA}

\author[0000-0001-5082-9536]{Yuan-Sen Ting(丁源森)}
\altaffiliation{Hubble fellow}
\affiliation{Institute for Advanced Study, Princeton, NJ 08540, USA}
\affiliation{Department of Astrophysical Sciences, Princeton University, Princeton, NJ 08544, USA}
\affiliation{Observatories of the Carnegie Institution of Washington, 813 Santa Barbara Street, Pasadena, CA 91101, USA}
\affiliation{Research School of Astronomy \& Astrophysics, Australian National
University, ACT 2611, Australia}

\author[0000-0003-4996-9069]{Hans-Walter Rix}
\affiliation{Max Planck Institute for Astronomy, K\"onigstuhl 17, D-69117 Heidelberg, Germany}

%%%%%%%%%%%%%%%%%%%%%%%%%%%%%%%%%%%%%%%%%%%%%%%%%%%%%%%%%
% Title 
%%%%%%%%%%%%%%%%%%%%%%%%%%%%%%%%%%%%%%%%%%%%%%%%%%%%%%%%%
\title{Keeping it Cool: Much Orbit Migration, yet Little Heating, in the Galactic Disk}

%%%%%%%%%%%%%%%%%%%%%%%%%%%%%%%%%%%%%%%%%%%%%%%%%%%%%%%%%
%
%
%
%                 Abstract & Keywords
%
%
%
%%%%%%%%%%%%%%%%%%%%%%%%%%%%%%%%%%%%%%%%%%%%%%%%%%%%%%%%%

\begin{abstract}
A star in the Milky Way's disk can now be at a Galactocentric radius quite distant from its birth radius for two reasons: either its orbit has become eccentric through radial heating, which increases its radial action $J_R$ (`blurring'); or merely its angular momentum $L_z$ has changed and thereby its guiding radius (`churning'). We know that radial orbit migration is strong in the Galactic low-$\alpha$ disk and set out to quantify the relative importance of these two effects, by devising and applying a parameterized model ($\ppm$) for the distribution $p(L_z, J_R, \tau, \fe | \ppm)$ in the stellar disk. This model describes the orbit evolution for stars of age $\tau$ and metallicity $\fe$, presuming coeval stars were initially born on (near-)circular orbits, and with a unique $\fe$ at a given birth angular momentum and age. 
We fit this model to APOGEE red clump stars, accounting for the complex selection function of the survey.
The best fit model implies changes of angular momentum of $\sqrt{\langle \Delta L_z \rangle^2} \approx 619\, \mathrm{kpc~km/s~}(\tau/\mathrm{6~Gyr})^{0.5}$, and changes of radial action as $\sqrt{\langle \Delta J_R \rangle^2} \approx 63\, \mathrm{kpc~km/s~} (\tau/\mathrm{6~Gyr})^{0.6}$ at 8 kpc. This suggests that the secular orbit evolution of the disk
is dominated by diffusion in angular momentum, with radial heating being an order of magnitude lower.
\end{abstract}


%%%%%%%%%%%%%%%%%%%%%%%%%%%%%%%%%%%%%%%%%%%%%%%%%%%%%%%%%
%
%
%
%                Section: Introduction
%
%
%
%%%%%%%%%%%%%%%%%%%%%%%%%%%%%%%%%%%%%%%%%%%%%%%%%%%%%%%%%

\section{Introduction}
\label{sec:introduction}

What dominates the secular orbit evolution of nearly isolated disk galaxies? The Milky Way's last major merger is thought to have occurred before 7-8 Gyr ago \citep[e.g.][]{rix_bovy_2013,bland-hawthorn_gerhard_2016,belokurov_etal_2018,Helmi_etal2018}, leaving a long subsequent period for internal processes to dominate the dynamical evolution of the disk: non-axisymmetries (bar, spiral arms) and giant molecular clouds can rearrange stellar orbits in different ways \citep[e.g.][]{sellwood2014, sellwood_binney_2002}, as they resonantly interact or scatter.

The change in a star's orbit can be decomposed into (1) `cold processes': the orbit's size or angular momentum changes, but remains circular, and (2) `heating processes': the orbit's eccentricity and vertical extent change. \cite{sellwood_binney_2002} dubbed the first process ``radial migration" and postulated that it could be important for restructuring stellar discs. Pure radial migration occurs when stars are near corotation of a non axisymmetry and can change angular momentum ($L_z$) by some amount $\Delta L_z$, without significant change in their radial action ($J_R$): $\Delta J_R = 0 \Delta L_z$. Such secular change of angular momentum can be very substantial -- at least in dynamical simulations -- indeed of order unity \citep{roskar_etal_2008a, minchev_2011, kubryk_etal_2013, halle_2015, loebman_etal_2016}. Nonetheless, \cite{minchev_2011} and \cite{daniel_2019} used simulations to argue that such interactions should come with non-zero changes in $J_R$ due to possible resonance overlaps between, for example, co-rotation and Lindblad resonances of different non-axisymmetries, or higher order resonances. But this effect could also increase the angular momentum changes. Heating processes could arise from interactions with non-axisymmetric perturbations through other resonances, or other heating agents such as satellites \citep[e.g.][]{VelazquezWhite_1999}. The different components of the velocity dispersion have been measured to increase with stellar ages both in the Solar neighbourhood and over the disk \citep{wielen_1977, soubiran_2008, sanders_das_2018, ting_rix_2019,  mackereth_2019}.

There is now persuasive evidence that {\it radius} migration in the Galactic disk is strong. In \cite{frankel_etal_2018} we measured this radius migration, which must be the combined effect of radial heating and angular momentum diffusion. We modeled radial migration as a global process, fitting the distributions $p(R|R_0, \tau)$ of present-day Galactocentric radii ($R$) as a function of their birth radii ($R_0$) and age ($\tau$) over a wide range of radii. We assumed that stars were born on initially tight metallicity-birth radius relations, and it is the the radial diffusion that introduces the observed present-day scatter in this relation \citep{edvardsson_etal_1993, casagrande_etal_2011}. Inferring radial migration this way was originally proposed in the seminal paper by 
\cite{schonrich_binney_2009a}. Our global model fits turned out to constrain the overall migration scale well, to $\sigma_{RM} = 3.6 \mathrm{kpc}\sqrt{\tau/\mathrm{8 ~Gyr}}$: the typical star migrates about by a scale-length over the age of the disk.

However, this work did not disentangle diffusion in angular momentum from the increase in radial action, but only measured the combined effect of `heating' and `cold $L_z$ diffusion' as `orbit migration'. Therefore, the relative contributions of $\Delta J_R$ \citep[or `blurring' in the terminology of][]{sellwood_binney_2002,schonrich_binney_2009a} and $\Delta L_z$ (or `churning') to the evolution of the stars' orbits in the Milky Way has yet to be quantified. This is what we set out to do here.

We now set out to build on that model and disentangle the strength of diffusion in angular momentum from that of increase in radial action. We will do this by generalizing the model of F18 from constraining the radius migration, i.e. the `diffusion rate' in $R$,  to constraining the secular orbit evolution in the disk plane, by quantifying the diffusion rates of both actions $L_z$ and $J_R$. In Section \ref{section_data}, we describe our data set. We then construct the model and data likelihood in sections \ref{sec:model} and \ref{sec:likelihood} and present the result best fit model in section \ref{sec:results}. We discuss the implications in Section \ref{sec:discussion}.

%%%%%%%%%%%%%%%%%%%%%%%%%%%%%%%%%%%%%%%%%%%%%%%%%%%%%%%%%
%
%
%
%          Section: Data
%
%
%
%%%%%%%%%%%%%%%%%%%%%%%%%%%%%%%%%%%%%%%%%%%%%%%%%%%%%%%%

\begin{figure*}
    \centering
    \includegraphics[width=\textwidth]{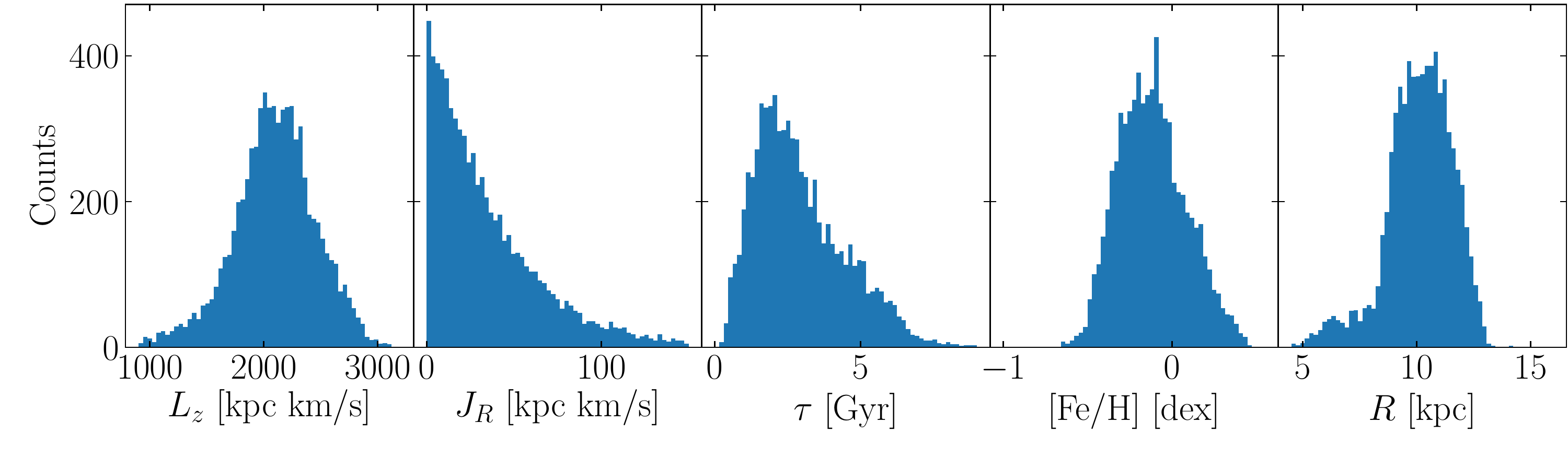}
    \caption{Distributions of the adopted sample of red clump stars (drawn from APOGEE DR14, low-$\alpha$) in angular momentum ($L_z$), radial action ($J_R$), age ($\tau$), $\fe$, and Galactocentric radius ($R$). The $L_z$ and $R$ distributions are clearly affected by spatial selection effects due to APOGEE's footprint. The sample's age distribution, with a prominent peak at 2 Gyr, reflects the combination of the `underlying' age distribution and (most prominently) the age- or mass-dependent time duration of the core helium burning evolutionary stage (red clump).}
    \label{fig:distributions_data}
\end{figure*}

\section{Data: APOGEE-DR14 Red Clump Giants}
\label{section_data}

Since the disk build up and its secular dynamical evolution involve processes occurring on large spatial and time scales, our science case requires a sample of stars with extensive coverage of the Galactic disk, a wide range of ages, and accurate and reliable 6D phase-space coordinates to calculate orbits. In practice, this requires spectroscopic and astrometric information, precise distances (as these are the dominant uncertainty in the action determination), and a way to limit the impact of dust extinction at low Galactic latitudes. A crossmatch of the APOGEE-RC catalogue \citep{bovy_etal_2014,majewski_etal_2017,Abolfathi_2018_apogeedr14} with Gaia DR2 \citep{gaia_dr2_2018, Lindegren_2018_gaiadr2} satisfies these desiderata: red clump stars are bright standard candles and APOGEE collected their spectra in near-infrared (limiting the effect of extinction).

\subsection{Data Selection and Catalogs}
We start with the stars from the 14th data release of the APOGEE near-infrared spectroscopic survey and restrict to the low-$\alpha$ sample as in \cite{frankel_etal_2018}, with stars mostly born between $\sim$ 8 Gyr ago and now. %\citep{Abolfathi_2018_apogeedr14, majewski_etal_2017, bovy_etal_2014}
We further select high-fidelity red clump stars derived by \cite{ting_hawkins_rix_2018} cross-matched with Gaia DR2. 

Red clump stars are low-mass stars in the core helium burning stage. They are good standard candles \citep[to $\sim 0.1$~mag,][]{girardi_2016, hawkins_etal_2017,hall_etal_2019, sanders_smith_evans_2019,chan_bovy_2019}, allowing the determination of a precise photometric distance. The identification of the red clump stage, done in \cite{ting_hawkins_rix_2018}, relies on spectroscopic estimates of the asteroseismic parameters $\Delta \nu$ and $\Delta P$ which contain information on the evolutionary stage and stellar mass. 

We further restrict our sample to the low latitude ($|b|< 25$ deg) ``short cohort'' \citep[as defined in][]{zasowski_2013} fields of APOGEE, which consists of the brightest stars with $H$  band apparent magnitude $7 \leq H \leq 12$ to ease our modeling and reduce the fraction of stars with large uncertainties. Stars in longer cohorts are fainter, so at larger distances, have greater distance uncertainties and are more extinguished. Including them would increase computational expenses and complicate the selection function without commensurately increasing the information content. So we restrict the analysis to the short cohort stars for which we can work out the probability that they were selected for targeting in APOGEE ({\it i.e.} we can determine the selection function). We reject all APOGEE ``special targets'' in our data set as well as all the stars that are in APOGEE fields for which we could not work out a probability of selection \citep[see][]{frankel_etal_2019}. Since the number of crossmatch failures between the red clump sample and Gaia DR2 is negligible ($ < 1\%$), we assume in the following that the selection of our data is purely determined by APOGEE selection function, and that Gaia is complete within the APOGEE's short cohort selection cuts.

\vspace{9mm}
\subsection{Basic Data and their Uncertainties}
We use eight basic pieces of information: Galactic longitude ($l$), Galactic latitude ($b$), distance ($D$), line of sight velocity ($v_\mathrm{los}$), metallicity (\fe) from APOGEE, proper motion in right ascension ($\mu_\alpha$) and declination ($\mu_\delta$) from Gaia, and age ($\tau$) derived from the full spectrum in \cite{ting_rix_2019}, which are calibrated to asteroseismology.

The photometric distances ($D$) were determined to about 7\% in \cite{ting_rix_2019}, using near-infrared and Gaia G photometry for red clump stars as standard candles, exploiting the fact that interstellar extinction is weaker at longer wavelengths \citep[e.g.,][]{indebetouw_2005, wang_2019_extinction}.
The line of sight velocity ($v_\mathrm{los}$) is taken from the APOGEE-DR14 catalog,
and the proper motions ($\mu_\alpha, \mu_\delta$) from Gaia DR2.
Spectroscopic age estimates $\tau$ for this sample were obtained by \cite{ting_rix_2019} from a data-driven method built to determine ages
from the APOGEE spectra, trained on the APOKASC2 red clump sample \citep{pinsonneault_2018}; this approach has a precision of about 0.15 dex, with possible systematics at large ages, because the C and N spectral features tracing the age, \citep[e.g.][]{Martig2016,ness_etal_2016} vary more weakly at large ages and hence contains less information. A more extensive discussion on the possible implications of such systematics on the modeling, and comparisons of different age estimates can be found in \cite{frankel_etal_2019}.
The metallicity estimates (\fe ) are taken from the ASPCAP pipeline with typical uncertainties below 0.05~dex \citep{holtzman_2018A_dr14}.

\subsection{Galactocentric Rest Frame and Orbital Actions}
From these basic data, we extract and pre-compute the quantities that are more directly used in our model, and propagate uncertainties via Monte Carlo sampling of 80 points. In particular, we compute the Galactocentric radius ($R$), height above the plane ($z$), azimuthal velocity ($v_\phi$) and radial velocity ($v_R$) in Galactocentric coordinates, assuming the distance between the Sun and the Galactic center $R_\odot = 8.2 $ kpc \citep{gravity_2019}, and the Sun's height above the Galacctic plane $z_\odot \approx 20.8$ pc \citep{bennett_bovy_2019}. We also assume the Solar velocity with respect to the Local Standard of Rest $v_\odot = [-11.1, 12.24,~7.25]~\mathrm{km.s^{-1}}$\citep{schoenrich_2010} and the tangential velocity of the Sun 247.4 km/s \citep{gravity_2019, reid_2004}.

We further compute the orbital parameters relevant to our modeling of the Galactic disk; going from phase-space coordinates $(\vec{x},\vec{v})$ to orbits, as quantified {\it e.g. } by their actions, requires the adoption of a gravitational potential.  We compute the stars' angular momenta ($L_z$, which are the azimuthal actions $J_\phi$ in an axissymmetric potential) and radial actions ($J_R$) using the Python package {\sl Galpy} \citep{bovy_2015_galpy} based on the algorithm of \cite{binney_2012}. The orbits are integrated over the potential {\sl MWPotential2014} of this package, which is also the potential used in our model, scaled to the Galactocentric rest-frame as described above. The 1D marginal distributions of angular momentum, radial action, age, \fe~ and Galactocentric radius are illustrated in Figure \ref{fig:distributions_data}. The distributions in angular momentum and Galactocentric radius are directly affected by the APOGEE on-sky footprint and implicit distance limit; else we would expect these distributions to be approximately exponential. The age distribution shows a prominent peak at at $\sim$2 Gyr, reflecting the mass dependence of the RC lifetime \citep{girardi_2016, bovy_etal_2014}.

%%%%%%%%%%%%%%%%%%%%%%%%%%%%%%%%%%%%%%%%%%%%%%%%%%%%%%%%%
%
%
%
%          Section: Methodology
%
%
%
%%%%%%%%%%%%%%%%%%%%%%%%%%%%%%%%%%%%%%%%%%%%%%%%%%%%%%%%%
\vspace{7mm}
\section{Chemo-dynamical Model for the Evolution of the Galaxy's Low-$\alpha$ Disk \label{sec:model}}

%-----------------------------------------
%
% Graphical model
% 
% The caption is in the file
% graphical_model.tex 
%
%------------------------------------------
%%%%%%%%%%%%%%%%%%%%%%%%%%%%%%%%%%%%%%%%%%%%%
%
% The caption of this graphical model
% is at the bottom of the page
%
%%%%%%%%%%%%%%%%%%%%%%%%%%%%%%%%%%%%%%%%%%%%%%

\usetikzlibrary{shapes,arrows,positioning,fit}

\begin{figure*}
\centering
%
% Start the graphical model
%
\begin{tikzpicture}[->,>=stealth']

% define colors
\definecolor{bblue}{RGB}{34,94,168}
\definecolor{ggreen}{RGB}{35,132,67}
\definecolor{rm}{RGB}{140,45,4}

%---------------------------------------
% Make some commands
%---------------------------------------
\tikzstyle{main}=[circle, minimum size = 12mm, line width=0.4mm, draw =black!80, node distance = 7mm]
\tikzstyle{ellip}=[ellipse, minimum width=20mm, minimum height=12mm), line width=0.4mm, draw =black!80, node distance = 7mm]
\tikzstyle{err}=[circle, minimum size = 2mm, thick, draw =black!100, node distance = 9mm]
\tikzstyle{connect}=[-latex, thick]
\tikzstyle{box}=[rectangle, draw=black!100]
\tikzstyle{line}=[draw]

%-----------------------------------------
% Make Nodes
%-----------------------------------------
   % 0th floor = observables
   \node[ellip, minimum width=7mm, fill = black!15] (xv) [label={[align=center]center:$R_i$\\ $z_i$}] { };
   \node[ellip, minimum width=7mm, fill=black!15] (v) [left=of xv][label={[align=center]center:$v_{Ri}$\\ $v_{\phi i}$}] { };
   \node[ellip, minimum width=15mm,fill = black!15] (feobs) [right=of xv, label=center:{[Fe/H]$_{\mathrm{obs}i}$}] { };
   \node[main, fill = black!15] (tobs) [right=of feobs, label=center:$\tau_{\mathrm{obs}i}$] { };
   \node[ellip, minimum width=28mm, fill = black!15] (evol) [right=of tobs][draw, align=center] {Evolutionary  \\stage $i$ };
   
   % 1st floor: action and metallicity
   \node[main, fill=white!100](J)[above=of xv, label=center:{$\boldsymbol{J}_i, \boldsymbol{\theta}_i$}] {};
   \node[main, fill=white!100](fe)[above=of feobs, label=center:{[Fe/H]$_i$}] {};
   
   % 2nd floor: birth conditions: birth angular momentum, age
   \node[main, fill=white!100](Lz0)[above=of J, label=center:{$L_{z0i}$}] {};
   \node[main, fill=white!100](t)[above=26mm of tobs, label=center:{$\tau_i$}] {};

   % 3rd floor: birth parameters
   \node[ellip, fill=white!100](Rd0)[above=of Lz0][draw=ggreen, line width=0.8mm, align=center]{Formation\\ structure, $R_{d0}$};   
 %  \node[ellip, fill = white!100](SFH)[above=of t, label=right:{Star formation history $\tau_\mathrm{SFR}, x$}][draw=ggreen, line width=0.8mm] {};
   \node[ellip, fill=white!100](SFH)[above=of t][draw=ggreen, line width=0.8mm, align=center]{Star formation \\history, $\tau_\mathrm{SFR}, x_{io}$};

   % Left:  Potential and secular evolution
   \node[ellip, minimum width=25mm, fill=white!100](secular)[left=20mm of J][draw=rm, align=center, line width=0.8mm]{Secular \\ evolution\\$\sigma_{Lz}$, $\sigma_{vR0}$, \\$R_{\sigma_R}$, $\beta$}; 
   \node[err, fill=black!100](vert)[left=20mm of Lz0, label={[align=center]left:Vertical\\ heating}]{};  
   %\node[err, fill=black!100](pot)[left=20mm of Lz0, label=left:$\Phi_\mathrm{pot}$] {};   
   
   % Right: chemical enrichment
   \node[ellip, minimum width=25mm, fill=white!100, draw=bblue, line width=0.8mm](chemical)[right=65mm of fe][draw, align=center]{Chemical \\ enrichment \\$\gamma_\mathrm{[Fe/H]}$ \\ $\mathrm{[Fe/H]_{max}}$ \\ $\mathrm{\nabla\mathrm{[Fe/H]}}$};
   
   % errors
   \node[err, fill=black!100](errdyn)[left=of v, label=below:$\sigma_{v}$] {}; 
   \node[err, fill=black!100](errfe)[below=of feobs, label=below:$\sigma_{[\mathrm{Fe}/\mathrm{H}]}$] {}; 
   \node[err, fill=black!100](errt)[below=of tobs, label=below:$\sigma_{\tau}$] {};
   
   % selection function
   \node[err, fill=black!100](selfunc)[below=of xv, label={[align=center]below: APOGEE \\spatial selection}]{};
   \node[err, fill=black!100](RC)[below=of evol, label={[align=center]below: Red Clump\\ selection}]{};
   
   % Potential
   \node[err, fill=black!100](pot)[left=11mm of selfunc, label=left:$\Phi_\mathrm{pot}$] {};

   % Rectangle around single objects to take the product over
   \node[rectangle, fit=(v) (Lz0)(evol), label=above right:{$ i=1,...,N $}] (rect) {};
   \node[rectangle, fit=(rect), inner sep=4.5mm, draw=black!100] {};

%------------------------------------------------
% Draw the connections
%------------------------------------------------
  \path (Rd0) edge [connect] (Lz0)
		(Lz0) edge [connect] (t)
		(SFH) edge [connect] (t)
		(pot) edge [connect] (v)
		(pot) edge [connect] (xv)
		(Lz0) edge [connect] (J)
		(Lz0) edge [connect] (fe)
		(t) edge [connect] (J)
		(t) edge [connect] (fe)
		(t) edge [connect] (evol)
		(t) edge [connect] (tobs)
		(secular) edge [connect] (J)
		(vert) edge [connect] (J)
		(J) edge [connect] (xv)
		(J) edge [connect] (v)
		%(xv) edge[bend right] node [left] {} (v)
		%(v) edge[bend left=-20] node [right] {} (xv)
		%(xv) edge [connect] (v)
		(fe) edge [connect] (feobs)
		(chemical) edge [connect] (fe)
		(errdyn) edge [connect] (v)
		(errfe) edge [connect] (feobs)
		(errt) edge [connect] (tobs)
		(selfunc) edge [connect] (xv)
		(RC) edge [connect] (evol);
\end{tikzpicture}

%%%%%%%%%%%%%%%%%%%%%%%%%%%%%%%%%%%%%%%%%
%% CAPTION
%%%%%%%%%%%%%%%%%%%%%%%%%%%%%%%%%%%%%%%%%
\caption{\label{fig_graphical_model} Probabilistic graphical model for the joint distribution $p(\mathbf{x_i}, \mathbf{v_i}, \fe_i, \tau_\mathrm{obsi}\} ~|~ \ppm)$ 
of the APOGEE data set. This simplified model reflects the combination of a global model for the Milky Way disk,  APOGEE selection function, and the marginalization over the data uncertainties. 
The circles filled in gray contain the APOGEE red clump stars' observables to be modelled. 
The circles inside the large contain the variables in which model is cast : the true birth angular momentum, true age, true action vector, true metallicity. 
The ellipses outside the rectangle contain the global model parameters to be fit: scale length at birth, star formation time-scale, inside-out parameter, 
secular evolution parameters and chemical enrichment parameters. The black points are fixed aspects of the model: present-day potential of the Milky Way, 
noise model and APOGEE selection function.}
\end{figure*}
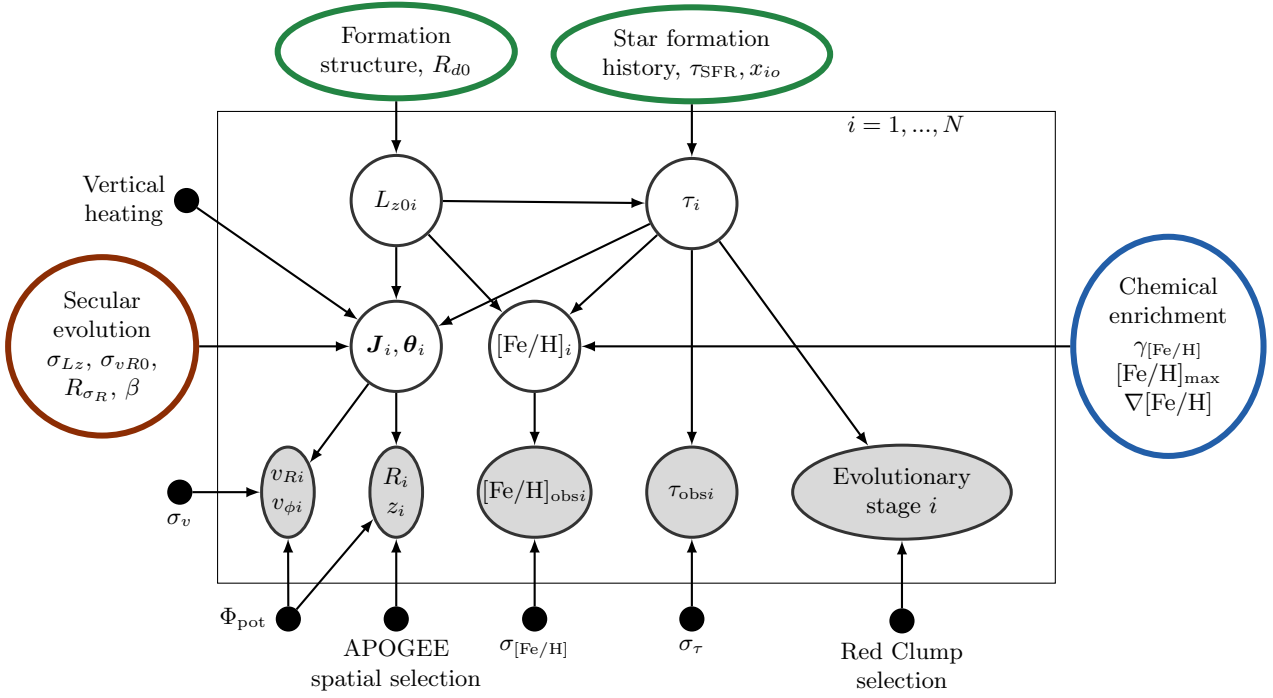

We now present a global model for the formation and evolution of the Galaxy's low-$\alpha$ disk, building directly on \cite{sanders_binney_2015}, \cite{ frankel_etal_2018} and \cite{frankel_etal_2019}. The most important astrophysical assumption of the model is that stars form on near-circular orbits from chemically well mixed cold gas with an inside-out star formation history. Over time, the orbits of stars evolve, and the gas is enriched in metals. The combination of age and \fe~in this model implies a birth angular momentum \Lzo~ or a birth radius. {\sl Radial heating} of stellar orbits is modeled as an increase of their mean radial action $J_R$ and {\sl radial migration} is modeled as a global diffusion process in angular momentum $L_z$. In our modeling, vertical heating is only implicit: we ignore the (weak) coupling between in-plane and vertical motions, and model an age-dependent vertical profile for the disk to incorporate the 3D spatial selection function (see Section \ref{sec:likelihood}). The overall model for the dataset $p_\mathrm{dataset}(l, b, D, \fe, \tau, v_R, v_\phi|\ppm)$, and how it is combined with the Galactic disk model $p_{MW}(L_z, J_R\fe, \tau |\ppm)$ is summarized in Figure \ref{fig_graphical_model} and in Table \ref{table:model_aspects}.
The model aspects are then combined together in the following section in a likelihood function, used to constrain its parameters.
In this way we can disentangle the strength of radial migration and radial heating and compare them, but only under a set of (physically sensible) assumptions that we will now lay out in some detail.
%---------------------------------------------------------------------
% Subsection: Assumptions
%---------------------------------------------------------------------
\subsection{Model Assumptions \label{section_assumptions}} 
Our modelling assumptions are as follows:
\begin{enumerate}
\setlength{\itemsep}{0pt}
\item Stars are born with a tight relation between \fe~and \Lzo~at any given $\tau$. This is a `weak chemical tagging' assumption, supported by \cite{krumholz_ting_2018} and  \cite{ness_etal_2019}, who showed that in the low-$\alpha$ disk, \fe and age can predict present-day $L_z$ precisely (and we here attribute the scatter in that relation to radial migration). \update{This assumption is also well supported by extragalactic obervations of a sample of spiral galaxies, finding that the azimuthal variations and scatter in the interstellar medium's [O/H] is low \cite[$<0.05$ dex in e.g.,][]{kreckel_2019}}. We parametrize this tight relation (Section \ref{section_enrichment}) and fit for the parameters.

\item Secular processes (processes happening on time scales longer than a typical orbital time scale) have dominated the evolution of the Galaxy's low-$\alpha$ disk. As the last major merger presumably occurred before the formation of the low-$\alpha$ disk \citep[e.g.,][]{rix_bovy_2013, bland-hawthorn_gerhard_2016,belokurov_etal_2018,Helmi_etal2018}, the Milky Way is thought to have evolved in near isolation (i.e. interacting mainly with itself and occasionally with lower mass satellites, e.g., the Sagittarius dwarf galaxy) for the past 7-8 Gyr, which leaves a lot of time for slower, more gradual processes to occur and affect the Galaxy's evolution. We assume that secular processes cause orbits to diffuse, and set out to measure the strength of this diffusion, but we do not make assumptions on the nature of this process nor try to identify the agents driving them.

\item We assume that the present-day potential of the Milky Way is sufficiently well approximated by the axisymmetric potential {\sl MWPotential2014} in the {\sl Galpy} package \citep{bovy_2015_galpy}, and that the present equilibrium state of the Milky Way disk can be described by separable distribution functions $p({\bf J}) = p(L_z)p(J_R|L_z)p(J_z|L_z)$ where  the vertical motion is independent of the radial action \citep[the so-called adiabatic approximation,][]{binney_2010}. This is manifestly an approximation since we know the disk is not axisymmetric, nor in equilibrium: there are spiral perturbations, a bar and a warp \citep{bland-hawthorn_gerhard_2016,beane_etal_2019}. 
\item When computing the actions, we use the St\"ackel approximation \citep{binney_2012} as implemented in the {\sl Galpy} package \citep{bovy_2015_galpy}.
Furthermore, we implicitly assume in the models that the height above the plane $z$ and vertical velocity $v_z$ are independent of the radial action \citep[the adiabatic approximation,][]{binney_2010}. 
We use this approximation because we wish to focus on the in-plane distributions $p(J_R|L_z)p(L_z)$ and for our considered orbits, which predominately lie close to the Galactic plane, the approximation is valid. 
\end{enumerate}

%---------------------------------------------------
% subsection: Build up of the disk
%---------------------------------------------------
\subsection{Modeling the Gradual Build-up of the Stellar Disk \label{subsection:pLzo}}
We now describe the parameterized version of our modeling for the successive
build-up of the Galactic stellar disk, encompassing the star-formation history with inside-out growth.

We model the time-integrated distribution of angular momenta at birth as
\begin{equation}\label{eq:Lz0_distribution}
    p(\Lzo|\ppm) = \frac{\Lzo}{\langle \Lzo \rangle ^2} \exp\left(-\frac{\Lzo}{\langle \Lzo \rangle}\right),
\end{equation}
where we define the mean angular momentum at birth as $\langle \Lzo \rangle = \rdo \times 235 \mathrm{km.s^{-1}}$, to fit for and interpret the parameter $\rdo$ as a global scale-length. This scale-length is time-integrated and reflects the global profile of the disk after all stars are born.
For a cold disk at birth, this model is approximately equivalent to an exponential surface density profile ($\Sigma(R_0) \propto \exp(-\Ro / \rdo)$) with a scale-length $\rdo$.
But the actual spatial scale-length of the disk at birth may be different since the spatial distribution of stars will depend on gradual changes of the potential due to the on-going build up of the disk.
The possible inside-out growth (illustrated in Fig. \ref{fig:model_aspects}) is modeled through an $\Lzo$-dependent star formation, where the star formation time-scale depends linearly on birth angular momentum, causing the inner disk to form stars on shorter time-scales than the outer disk. The star formation history is adapted from \cite{frankel_etal_2019}, but now taken to be a function of birth angular momentum (rather than birth radius):
\begin{equation}\label{eq:SFH_def}
    \begin{split}
            \mathrm{SFH}(&\tau ~|~\Lzo, \ppm) = c(\Lzo, \ppm) \\ 
            &\times \exp\left[\frac{1}{\tsfr}\left((1-\xio \frac{\Lzo /235\mathrm{~km\,s}^{-1}}{\mathrm{8.2~kpc}})\tau - \tau_m \right) \right].
    \end{split}
\end{equation}
Here, $c(\Lzo, \ppm)$ is a normalization constant such that $\int \mathrm{SFH}d\tau = 1$ at any given \Lzo.
Since the star formation history is not the primary focus of this work, we treat it as a nuisance aspect of the model. We fit for and marginalize over the parameters $\xio$ and \tsfr. The parameter $\tau_m$ corresponds to the maximum stellar age we consider in the low-$\alpha$ disk, fixed to 6 Gyr.

%---------------------------------------------------
% subsection: Potential
%---------------------------------------------------
\vspace{4mm}
\subsection{Present-day Gravitational Potential \label{subsection:gravitational_potential}}

We assume that the present-day gravitational potential of the Milky Way disk is well described by the {\sl MWPotential2014} in the {\sl Galpy} Package. In the present work, we only use the present-day potential of the Milky Way disk and its derived quantities (circular velocity $v_\mathrm{circ}$, epicyclic frequencies $\kappa$ and $\nu$ etc.) and make no assumptions about its past evolution.

As the Milky Way's stellar disk has gradually grown from inside-out over the past 7-8 Gyr (Subsection \ref{subsection:pLzo}), the mass distribution of the stellar and gas disks have changed, and the potential $\Phi_\mathrm{pot}(\tau)$ has evolved accordingly. Linking stellar birth radii $\Ro$ to their birth angular momenta $\Lzo$ would require to know exactly how the potential has evolved. This could be done by modeling the mass distribution in different Milky Way components, including a growth for the stellar disk. But the present-day contributions of each component of the Milky Way are already fairly unconstrained today \citep[e.g.,][]{de_salas_2019, eilers_etal_2019}, so we do not attempt to infer them in the past. Instead, our modeling is fully based on linking birth actions to present-day actions rather than birth positions. In an axisymmetric potential, if the gradual build up of the disk is slow and adiabatic, then the actions of stars should be conserved: stars on the same orbit but at different phases will, if the change in potential is sufficiently slow, experience the same changes of potential (averaged over a period) and thus conserve their actions. However, stellar velocities and positions should change: as the disk mass increases, stars will on average sink to orbits closer to the Galactic center. Therefore, measuring a change of orbital action $\Delta \mathrm{J}$ bypasses other orbital changes of the stars and gives direct insights into the secular processes in the disk.

%---------------------------------------------------
% subsection: Churning
%---------------------------------------------------
\subsection{Modeling the Angular Momentum Evolution \label{subsection:p_Lz}}

We follow the argument of \cite{sellwood_binney_2002} that the radial orbit redistribution of stars is caused by a sequence of stochastic processes, of some nature that we do not determine (which could be, for example, short-lived spiral perturbations). In this limit, stars follow a random walk in angular momentum so that radial migration can be modeled as a diffusion process in angular momentum, dubbed ``churning'' by \cite{sellwood_binney_2002}. 
Following \cite{sanders_binney_2015} we adopt the parameterized 
angular momentum 
diffusion equation
\begin{equation} \label{eq:diffusion}
    \frac{\partial f}{\partial t} = \frac{\partial}{\partial L_z}\left(-D^{(1)}
    f+ \frac{\sigma ^2}{2}\frac{\partial  f}{\partial L_z } \right).
\end{equation}
The nature of the radial migration process determines the diffusion coefficients of this diffusion equation. For simplicity, we assume that the diffusion coefficients in the equation above do not depend on $L_z$ and that $\sigma$ is independent of time. These assumptions imply that we are only constraining an effective global mean (in radius and time) of the overall $L_z$ diffusion.

\cite{herpich_etal_2017} showed that if radial migration were asymptotically efficient, the angular momentum distribution in a disk should go as $f(L_z) = \exp(-L_z/\langle L_z \rangle)/\langle L_z \rangle$, and their model predictions match the observed angular momentum profile of external galaxies well. On the other hand, surface brightness profiles of disk galaxies have been observed to have exponential or de Vaucouleur profiles \citep{deVaucouleur_1948}, which would correspond to (in the limit of circular orbits) an angular momentum distribution $f(L_z) \propto L_z  \exp(-L_z/\langle L_z \rangle)$ to which a bulge could be added at the center. Since we are not modeling a bulge in the present work, our model will in any case be inadequate in the inner few kpc of the Milky Way.

Drawing on these considerations, we impose that the steady state solution for the $L_z$ diffusion equation is either the exponential distribution in $L_z$, or the exponential surface density profile. At $L_z >> \langle L_z \rangle$\footnote{The region $L_z \leq \langle L_z \rangle $ corresponds to the inner 3~kpc of the disk, where we currently do not have data, see the right most panel of Fig.~\ref{fig:distributions_data}}, these distribution are similar and constrain the diffusion coefficient $D^{(1)}$ to $D^{(1)} =  \frac{-\srm ^2}{2 \langle \Lzo \rangle \tau_m}$, where we pose the mean specific angular momentum $\langle \Lzo \rangle = 235 \mathrm{km.s^{-1}} \rdo$. This ensures approximate conservation of angular momentum. In the microscopic limit, \cite{schonrich_binney_2009a} model this process with each star having a probability to move to a radius $r_i$ proportional to the stellar mass $m_i$ at that radius. This ensures that the number of stars migrating from radius $i$ to radius $j$ is proportional to $m_i m_j$, which is equal to the number of stars migrating from $j$ to $i$, thereby conserving the total disk profile. However, in reality the total disk angular momentum need not be conserved, as external torques, e.g. exchanges with the halo, could change it; here we do neglect this effect, since it is not well quantified or understood even in simulations: \cite{buck_2019} show that some Milky Way-like simulated galaxies see their scale-length increase over time (net outward migration), while others don't. Here, with our fixed choice of $D^{(1)}$ disk profile remains constant with a mean angular momentum $\langle \Lzo \rangle$. But individual populations of age $\tau$ born on profiles with $\langle \Lzo \rangle (\tau) < \langle \Lzo \rangle$ will, on average, broaden, and those born with  $\langle \Lzo \rangle (\tau) > \langle \Lzo \rangle$ will, on average, shrink.

The present day angular momentum of a star of age $\tau$ would then be related to its birth angular momentum by
\begin{equation}
\label{eq:p_radial_migration}
\begin{split}
p(\Lz ~|~ & \Lzo, \tau,~ \ppm) = N(\ppm)\\
& \times \exp{\left(-\frac{(\Lz - \Lzo - D^{(1)}\tau)^2}{2\sigma^2(\tau)}\right)}
\end{split}
\end{equation}
where $\sigma(\tau) = \srm\sqrt{\frac{\tau}{\tau_m}}$ is the radial migration strength in angular momentum units (kpc~km/s$^{-1}$). The parameter $\srm$ is to be fit and $N(\ppm)$ is a normalizing constant such that $\int_0 ^\infty p(\Lz ~|~ \Lzo, \tau, \ppm) d L_z= 1$ (there are no counter rotating stars in the disk). $\tau_m$ is the maximum age of the disk we consider, which is fixed to 6 Gyr here since there are so few constraining stars in the data at larger ages (see 3rd panel in Fig. \ref{fig:distributions_data}).

%---------------------------------------------------
% subsection: blurring
%---------------------------------------------------
\subsection{Radial Heating \label{subsection:p_heating}}

We assume that stars are born on near-circular orbits, and that their mean radial action increases with time as their orbits are kinematically heated (dubbed ``blurring'' by \cite{sellwood_binney_2002}). We adopt the isothermal disk model from \cite{sanders_binney_2015, binney_2010}, and we fit for the increase of mean radial action as a function of age and final position. The radial action distribution is then written as
\begin{equation}\label{eq:p_radial_action}
p(J_R~|~\tau, \Lz,~ \ppm) = \frac{1}{2\pi}\frac{\kappa}{\sigma_R^2}
\exp\bigl (-\frac{\kappa J_R}{\sigma_R^2(\tau,\Lz)}\bigr ),
\end{equation}
with $\kappa = \kappa(R_\mathrm{circ}(\Lz))$ the frequency of radial motion (the epicycle frequency), which depends on the guiding radius $R_\mathrm{circ}$, the radius of circular orbit of angular momentum \Lz.
The velocity dispersion $\sigma_R = \sigma_R(\tau, R_\mathrm{circ}(\Lz))$ traces the heating history of stars in the disk:
\begin{equation}\label{eq:heating}
\begin{split}
    \sigma_R(\tau, R_\mathrm{circ}(\Lz)) &= \sigma_\mathrm{vR0} \left(\frac{\tau + \tau_1}{\tau_m + \tau_1} \right)^\beta\\ 
    & \times \exp\left(\frac{8\,\mathrm{kpc} - R_\mathrm{circ}}{R_{\sigma_R}}\right),
    \end{split}
\end{equation}
where $\sigma_\mathrm{vR0}$ is the velocity dispersion of stars of age $\tau_m = 6$ Gyr in the Solar neighbourhood (to be fitted), $\tau_1$ is set to reflect a dynamical time, $\sim$ 110 Myr, as in \cite{sanders_binney_2015} and allows stars to be born with small but non-zero eccentricity. The scale $R_{\sigma_R}$ represents a possible decay of the velocity dispersion with Galactocentric radius, as in \cite{sanders_binney_2015} (to be fitted).
Note again that the radial actions are adiabatic invariants, and so the gradual build up of the disk should not influence the radial action of stars after their birth.

%--------------------------------------------------
% Subsection: chemical enrichment
%--------------------------------------------------
\begin{figure*}
    \centering
    \includegraphics[scale=0.64]{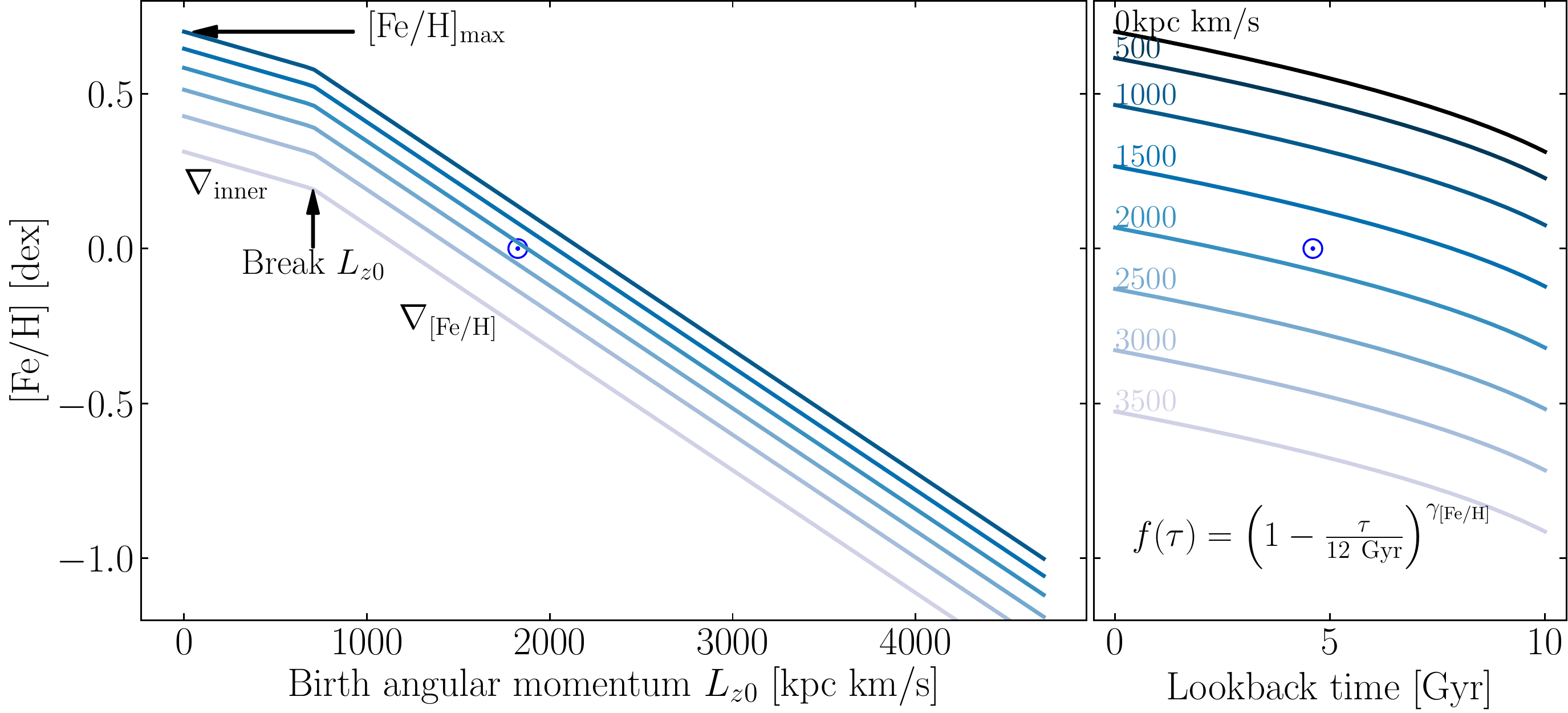}
    \caption{Functional forms to fit for $\fe (\Lzo, \tau, \ppm)$, showing \fe~as a function of \Lzo~in the left panel, and as a function of look-back time in the right panel (for a set of birth angular momenta). The model parameters chosen here are \gfe = 0.45, $\dfe = -0.093~\mathrm{dex.kpc^{-1}}$ and $\fe_\mathrm{max} = 0.70$ dex. The present-day gradient \dfe~sets the `outer' metallicity profile, the power index \gfe~sets the time dependence and $\fe_\mathrm{max}$ sets the present-day \fe~at the center of the Galaxy. We show the position of the Sun marked with an $\odot$ symbol, at $\fe_\odot=0$ dex and $\tau_\odot$ = 4.6 Gyr.}
    \label{fig:tagging_birth_feh}
\end{figure*}

\subsection{Weak Chemical Tagging: [Fe/H]-$\tau$-$\Lzo$ Relation \label{section_enrichment}}

For the current modeling, we use \fe, and assume that it is approximately only a function of time and $\Lzo$; in some sense, this is the weakest form of ``chemical tagging''.

We model the $\fe$-$\tau$-$\Lzo$ relation as a a power law in time and a broken line in the variable $\Lzo$ with an inner disk radial gradient of $-0.03~\mathrm{dex/kpc}$ for $\Lzo<3\times235\,\mathrm{kpc\,km/s}$. Since we have no constraining data on the inner metallicity profile in the disk, we use two external arguments to motivate this imposed flattening in the inner disk: (1) the observed metallicity profile in the stars is flatter in the inner disk/bulge  \citep{pietrukowicz_2015}, and (2) chemical evolution models predict a flatter metallicity profile in the gas in the inner disk \citep[e.g.,][]{schonrich_binney_2009a}. Namely, 
\begin{equation}\label{eq:eq_Fe_Ro_t}
\begin{split}
\fe = & \fe_\mathrm{max}f(\tau ) + b_\fe \\ 
&+ \dfe(\Lzo)  \frac{\Lzo}{235~\mathrm{kms^{-1}}}
\end{split}
\end{equation}
where the gradient
\begin{equation}
  \dfe(\Lzo) =\begin{cases}
    \nabla\mathrm{inner} &\text{$\frac{\Lzo}{235 \mathrm{km.s^{-1}}} < 3$ kpc}\\
    \dfe & \text{otherwise}.
  \end{cases}
\end{equation}
with the inner metallicity gradient $\nabla\mathrm{inner}$ fixed at $-0.03~\mathrm{dex.kpc^{-1}}$ and the outer metallicity gradient $\dfe$ is to be fitted. $b_\fe$ is introduced so the overall profile is continuous at $\Lzo = 3\times 235~\mathrm{kpc~km/s}$:
\begin{equation}
  b_\fe =\begin{cases}
     0 &\text{$\frac{\Lzo}{235 \mathrm{km.s^{-1}}} < 3$ kpc}\\
    \frac{(\nabla_\mathrm{inner}-\dfe) \Lzo}{235~\mathrm{km~s^{-1}}} & \text{otherwise}.
  \end{cases}
\end{equation}
The central metallicity is governed by the parameter $\fe_\mathrm{max}$ and we use a time dependence of
\begin{equation}\label{eq:enrichment_time}
f(\tau ) = \left( 1 - \frac{\tau}{12~\mathrm{Gyr}} \right)^\gfe.
\end{equation}
The set of model parameters we fit for are \dfe, \gfe, and $\fe_\mathrm{max}$. These functions are plotted in Fig.~\ref{fig:tagging_birth_feh}.

%---------------------------------------------------
% subsection: Vertical distribution
%---------------------------------------------------
\subsection{Vertical Distribution of Stars \label{subsection:p_z}}
The vertical distribution of stars in the Milky Way disk is not the primary focus of this work (since we are essentially interested in the in-plane motions), but we must model it because the spatial selection of our data is in three spatial dimensions and so we require a 3D model for the disk. 
For simplicity, we model the vertical distribution of stars in the disk with the best fit model of \cite{ting_rix_2019} in the regime of the isothermal disk:
\begin{equation}
\begin{split}
p(z~|~L_z, \Lzo, \tau, \ppm) &=  \frac{1}{2h_z(L_z, \Lzo, \tau)}\\ & \times \mathrm{sech}^2\left(\frac{z}{h_z(L_z, \Lzo, \tau)}\right),
\end{split}
\end{equation}
where the scale-height
\begin{equation}
h_z = a_z \sqrt{\frac{2 \overline{J_z}(L_z, \Lzo, \tau)}{\nu (R)}},
\end{equation}
with the vertical frequency $\nu$ defined such that $\nu ^2 = \frac{\partial^2\Phi}{\partial z^2}$. $\overline{J_z}(L_z, \Lzo, \tau)$ is the mean vertical action of stars of angular present momentum $L_z$, birth angular momentum $\Lzo$ and of age $\tau$. \cite{ting_rix_2019} studied the vertical heating history of the Galactic disk using an APOGEE red clump data set, and published an analytic fit for $ \overline{J_z}(L_z, \Lzo, \tau)$.
We model the age- and radial-dependent vertical distribution of red clump stars according to this form. Since the model of \cite{ting_rix_2019} used different age dependencies and birth radii scales than those in our work, we allow for an overall scaling of their heating law parametrized by $a_z$, which we anticipate will be near unity (see resulting fits in Table \ref{table:model_aspects}).

%===========================================================================
%
%
%  Table summarizing the model aspects and model parameters
%
%
%===========================================================================
\begin{table*}
\begin{center}
\begin{tabular}{ |p{2.5cm}|c|c|c| }
\hline
%-------------------T titles  ---------------------------------------------------------------------------------------------------------
 Model aspect & Functional Family & Model Parameters & Max. Likelihood Value \\ 
 \hline
% ----------------- First row -------------------------------------------------------------------------------------------------------
\multirow{2}{*}{\parbox{2.5cm}{Global structure at birth}}
 & { {$p(\Lzo | \ppm)$}} & {Birth scale-length} & \\
 & $\propto \Lzo \exp{(-\Lzo / \vcirc \rdo)} $  & {{$\rdo$}} & {2.8 kpc} \\
 \hline
 % ----------------- 2nd row -------------------------------------------------------------------------------------------------------
\multirow{3}{*}{\parbox{2.5cm}{Inside-out star formation history}} & { $ p(\tau | \Lzo, \ppm)$} & {Inside-out coefficient $\xio$} & 0.65\\
& {$ \propto \exp\left[\frac{1}{\tsfr}\left((1-\xio \frac{\Lzo /\vcirc}{\mathrm{8.2~kpc}})\tau - \tau_m \right) \right]$} & {{SFR time-scale $ \tsfr$} } & 1 Gyr \\
& & Maximum age $\tau_m$ & Fixed to 6 Gyr \\
 \hline
 % ----------------- 3rd row -------------------------------------------------------------------------------------------------------
 \multirow{2}{*}{\parbox{2.5cm}{Angular momentum diffusion}}
 & {$ p(L_z | \Lzo, \tau, \ppm)$} & {Diffusion strength} & {} \\
 & {$\propto \exp{\left(-\frac{(\Lz - \Lzo - D\tau)^2)}{2\srm^2\tau/\tau_m} \right)}$} & {\srm} & 582 kpc\,km/s\\
 \hline
 % ----------------- 4th row -------------------------------------------------------------------------------------------------------
 \multirow{3}{*}{Radial heating} & { $ p(J_R | \Lz, \tau,  \ppm)$} & {Time dependence $\beta$} & 0.3  \\
 & {$\propto \exp(-\kappa J_R / \sigma_R^2)$} &Velocity dispersion $\sigma_\mathrm{vR0}$ & 49 km/s \\
 &$    \sigma_R = \sigma_\mathrm{vR0} \left(\frac{\tau + \tau_1}{\tau_m + \tau_1} \right)^\beta \exp\left(\frac{8\,\mathrm{kpc} - R_c}{R_{\sigma_R}}\right)$ & Radial dependence $R_{\sigma_R}$& 19 kpc\\
 \hline
 \multirow{1}{*}{Vertical heating} & $ p(z | L_z, \Lzo, \tau, \ppm)$  & $a_z$ (scaling)& 1.16  \\
 \hline
 % ----------------- 5th row -------------------------------------------------------------------------------------------------------
 \multirow{3}{*}{\parbox{2.5cm}{Weak chemical tagging}} & $ p(\fe | \Lzo, \tau, \ppm)$ & Time dependence \gfe & 0.456 \\
 & {$ \fe = \fe_\mathrm{max} \left( 1 - \frac{\tau}{12~\mathrm{Gyr}} \right)^\gfe$} & {Radial gradient \dfe} & -0.0936 dex/kpc\\
 & {$ + \dfe(\Lzo)  \frac{\Lzo}{v_\mathrm{circ}(\Lzo)} + b_\fe$} & {Max \fe ~ $\fe_\mathrm{max}$} & 0.7 dex \\
 \hline
\end{tabular}
\end{center}
\caption{Summary of the main model aspects (described in Section \ref{sec:model}) and best-fit parameters \label{table:model_aspects}}
\end{table*}

%---------------------------------------------------
% subsection: modeling the dataset
%---------------------------------------------------
\section{
The Likelihood of the APOGEE-RC $\times$ Gaia DR2 Data \label{sec:likelihood}}
To determine how this model is constrained by our data, we must construct the data's likelihood for any given set of model parameters, and on this basis sample the parameter's posterior probability distribution, {\it pdf}. We now lay out how to implement this.

For each star, our `data' are
\begin{equation}
\mathcal{D} = \{l, b, D, \tau, \fe, v_R, v_\phi\}
\end{equation}
with their associated uncertainties (see Section \ref{section_data}). As noted in Section \ref{section_data}, these data are not sampled directly from the underlying Milky Way disk distribution: the sample very much reflects both the distribution of stars in the Milky Way $p_{MW}(\mathcal{D})$, and the selection process $S(l, b, D, \tau)\equiv p(\mathrm{select}| \mathcal{D})$.
%can be modelled as a inhomogeneous Poisson point process \citep[e.g.,][]{rix_bovy_2013, bovy_etal_2012} with a rate
Therefore, the probability distribution of the data given our model parameters is
\begin{equation}
\begin{split}
p_\mathrm{dataset}(\mathcal{D}~&|~ \ppm) = C\cdot p_\mathrm{MW}(\mathcal{D} ~|~\ppm)\\
&\times S(l,b,D)f_{RC}(\tau)
\end{split}
\end{equation} 
where $p_\mathrm{MW}(\mathcal{D} ~|~\ppm)$ is the model of the Galactic disk, which is a combination of the model aspects described above as detailed in Appendix \ref{sec:appendix_model}. $C$ is a normalizing term specified in Eq.~\ref{eq:survey_volume} and Section~\ref{sec:survey_volume}. $S(l,b,D) = p(\mathrm{select}|~l, b, D)$ is the selection function, or the probability that a star ends up in the catalog given its observable properties. The observables for each star are a combination of its intrinsic properties and its position and velocity with respect to us, and the evolutionary stage is determined using the spectrum \citep{ting_hawkins_rix_2018}. Typically the selection function is a strong function of apparent magnitude. In the case of standard candles where the apparent magnitude is a function of distance (and extinction) only, such as RC stars, the selection function essentially reduces to a function of $(l,b,D)$. Finally, the term $f_{RC}(\tau)$ is the probability of a star to be on the red clump evolutionary stage given its age $\tau$ \citep{bovy_etal_2014}.
%We will label this function as $S(l,b,D,\tau)$. It encodes the full probability of selection of the RC stars, which is a combination of the spatial selection function, denoted $S(l, b, D)$, and age-dependent population selection, written as $S(\tau)$ (or $f_{RC}(\tau)$) \citep{bovy_etal_2014}. 
% We account for both terms in the modeling.

We determine the spatial selection function of APOGEE following the methods laid out in \cite{rix_bovy_2013, bovy_etal_2014} and \cite{frankel_etal_2019}, and extend the methodology to the DR14 data which includes APOGEE-2.
% to include later data releases (DR14). 
The main difference in the APOGEE-2 target selection for the main disk fields with respect to APOGEE-1 is the inclusion of two color bins: blue ($(J-K_s) < 0.8$ mag) and red ($(J-K_s) > 0.8$ mag) \citep{zasowski_2017}. This makes the selection function more complex because the  fraction of selected stars of different stellar types differs between APOGEE fields. However, we are not affected by this complexity because we are working with RC stars, assumed perfect standard candles \citep[with $(J-K)_0 = 0.68$ mag,][]{hawkins_etal_2017} lying fully in the blue color bin, such that we need not account for the selection fractions of the red color bin.
In practice, the spatial selection function, $S(l,b,D)$, is a piece-wise function in each APOGEE field, such that for a field $i$ centred at $(l,b)_i$ it can be expressed solely as a function of distance $S_i(D)$. The dependence on distance is more complex than a constant between $D_{min}$ and $D_{max}$ set by the stars absolute magnitude and the survey's magnitude cuts, due to the 3D spatial distribution of dust, which limits the fraction of stars seen at a given absolute magnitude and distance. We model this using the 3D dust map of \cite{green_etal_2019}. A detailed description of how this is incorporated into the model can be found in \cite{frankel_etal_2019}\footnote{The selection function is published with a tutorial of its use at \url{https://github.com/NeigeF/apogee_selection_function}}.

%---------------------------------------------------
% subsection: survey volume
%---------------------------------------------------
\subsection{Normalization of the PDF: Survey Volume \label{sec:survey_volume} }

To make $p_\mathrm{dataset}$ a probability density function, it must be normalized by
\begin{equation} \label{eq:survey_volume}
C^{-1} 
= V_S(\ppm) 
= \int _\mathcal{D} p_\mathrm{MW}(\mathcal{D} ~|~\ppm) S(l,b,D) \mathrm{d}\mathcal{D}
\end{equation}
which is a 7-dimensional integral over all the physical properties of the data
We refer to $V_S$ as the `survey volume'.
Two of the integrals $(l, b)$ can be transformed into a sum over APOGEE fields, if we assume that the properties of stars in the sky within a single APOGEE pointing are uniform. This is a valid assumption as the APOGEE fields are typically 3 deg across. We compute $V_S$ in the next subsection, and assemble the data likelihood in Subsection~\ref{subsection:likelihood}.

Expanding Eq.~\ref{eq:survey_volume}, the survey volume can be written
\begin{equation}
\begin{split}
    V_S(\ppm) &= \sum_{\mathrm{field}~i} \int_\mathrm{Data~space} p_\mathrm{MW}(\tau, \Lzo, J_R, \Lz , z, ~|~\ppm)\\
    & \times S_i(D) f_{RC}(\tau) \Omega_i D^2  \mathrm{d}\Lzo \mathrm{d}D \mathrm{d}v_R
    \mathrm{d}v_\phi \mathrm{d}\tau,
\end{split}
\end{equation}
This integral is not straightforward to compute. The integrand is proportional to a probability density that we cannot fully compute, but that we can sample. Therefore, we compute this integral through iterative importance sampling (details in Appendix \ref{sec:appendix_Vs}). We start with a `blind' optimization of the data likelihood by normalizing the likelihood with samples generated from a model with initial plausible guesses for the model parameters $\mathbf{p_{m,\mathrm{prop}}}$. We then use these best fit parameters to generate new samples, which better reflect the shape of the likelihood and can be used normalize the likelihood for further calculations.

%----------------------------------------------------------------------------
% subsection: constructing the data likelihood function
%----------------------------------------------------------------------------

%
%
%
% Insert MCMC corner plot
%
%
%
\begin{figure*}
\centering
\includegraphics[scale=0.306]{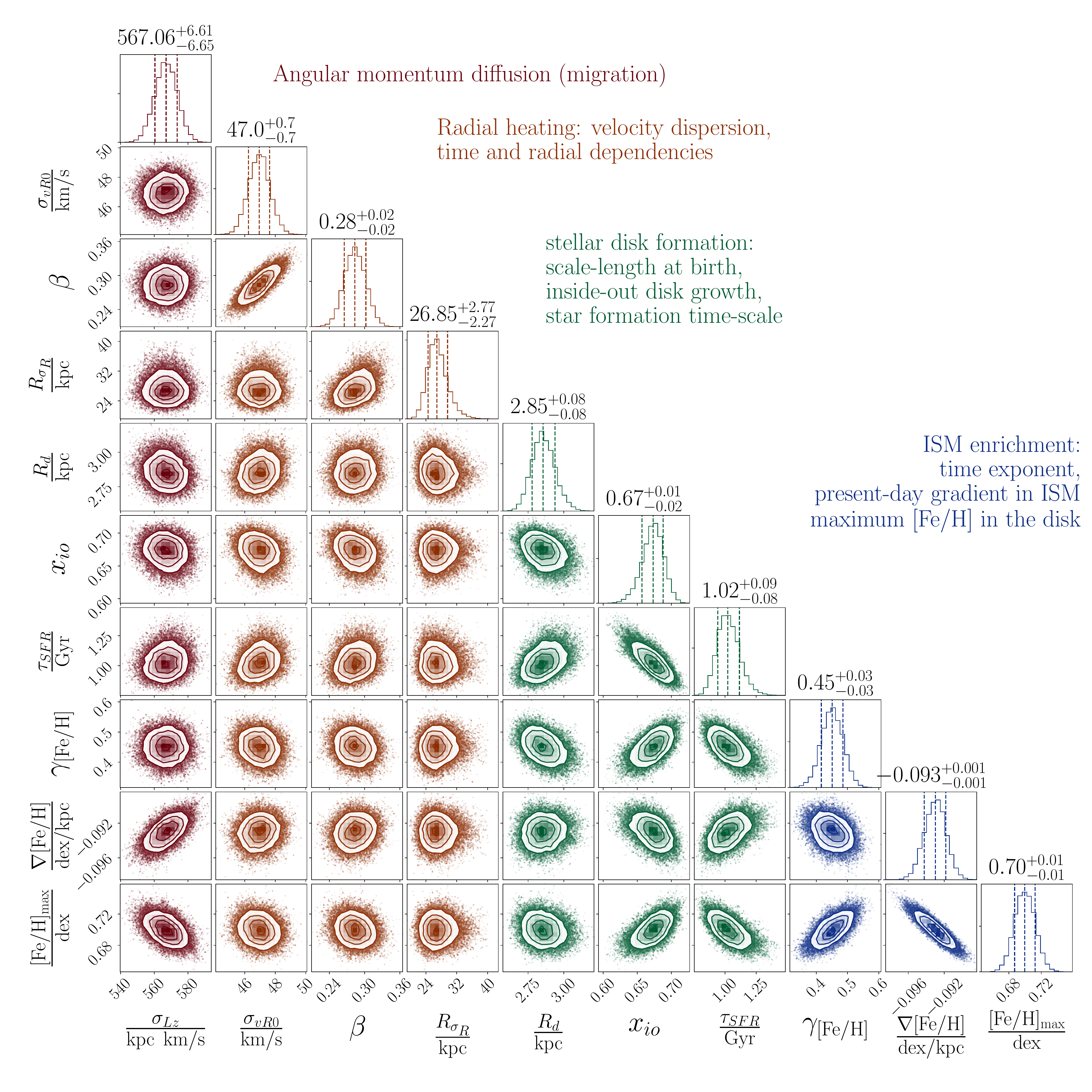}     
\caption{Posterior distribution of the parameters from a model fit to 7000 stars from APOGEE DR14. \srm~is the diffusion coefficient in angular momentum in kpc km/s. The other parameters are: radial velocity dispersion [km/s], time exponent for heating, scale-length of heating, disk scale length at birth, inside-out linear parameter, star formation time-scale, and enrichment parameters. Some parameters have important covariances, but all are well-constrained.\label{fig:results_corner}}
\end{figure*}

\subsection{Data Likelihood Function and Parameters Posterior\label{subsection:likelihood}}
The overall Milky Way disk model combined with the selection function 
predicts the likelihood of the data for any star in the sample.  Assuming all measurements are independent, we write the total likelihood of the entire data set given our model with parameters \ppm~as
\begin{equation} \label{eq:likelihood}
\begin{split}
    &p_\mathcal{L}( \{\fe, \tau, l,b,D,v_R, v_\phi\}|\ppm) \\
    &= \prod _{i=1}^{N_\mathrm{stars}} p_\mathrm{dataset}(\fe_i, \tau_i, l_i,b_i,D_i,v_{R,i}, v_{\phi,i}|\ppm).
\end{split}
\end{equation}
We use uniform priors with wide ranges in the parameter space, and enforce distances, spatial scales and time-scales to be positive. We first maximize the likelihood \citep{NeldMead65}, which gives the results in Table \ref{table:model_aspects}. We start the optimizer from different initial conditions to lower the chances of the optimizer becoming stuck in local maxima. Using the MCMC sampler {\sl emcee} \citep{foreman-mackey_2013}, we then sample the posterior, 
\begin{equation}
\begin{split}
    &p_\mathrm{pos}(\ppm~|~ \{\fe, \tau, l,b,D,v_R, v_\phi\})\\
    & \propto p_\mathrm{prior}(\ppm) p_\mathcal{L}(\{\fe, \tau, l,b,D,v_R, v_\phi\}~|~\ppm)
\end{split}
\end{equation}
using 12,000 iterations ($>$ 50 times the auto-correlation time) and 52 walkers. We initialize the MCMC sampling uniformly in a hypercube (of size greater than 8 times the error bars quoted in Fig \ref{fig:results_corner}) centered on the maximum likelihood estimates.

The posteriors of the model parameters are illustrated in Fig. \ref{fig:results_corner}.
The parameters of interest are the dynamical parameters: $\{\srm, \sigma_{vR0}, \beta, R_{\sigma_R} \}$, in red and orange in Fig.~\ref{fig:results_corner}. We treat the other model parameters as nuisance parameters and marginalize over them, although we comment briefly on them in Section~\ref{sec:results}.
There are no degeneracies and only weak covariances. In particular, the estimate of radial migration (or spread in angular momentum) \srm~is slightly correlated with the metallicity gradient. This is expected because the information on radial migration comes from the scatter in metallicity. Stars born in a galaxy with a shallower metallicity gradient will need to migrate larger distances to produce the same metallicity spread. The other notable covariance is between $\sigma_{vR0}$ and $\beta$, which together quantify heating in the radial direction: $\sigma_{vR0}$ is the strength, and $\beta$ the time dependency. If most of the sample is dominated by stars of a given age (here 2 Gyr), then the two ways to reproduce the local velocity dispersion of stars of that age is to either have a greater strength, and a faster time evolution (small $\beta$), or a smaller strength and a more linear time evolution (larger $\beta$).

The other model aspects also seem well fitted since the overall distributions of the data are well recovered by the model (Fig. \ref{fig:distributions_data}) and their estimate is robust to the tests we have performed (Subsection \ref{subsection:tests}). This shows that the best fit model is also qualitatively a good fit. The parameter analogous to the scale-length of the disk at birth is about 2.9 kpc. The covariances are more important for the inside-out growth model aspects, which are treated as nuisance parameters in the present work. The measurements of inside-out growth are consistent with our previous estimate in \cite{frankel_etal_2019} with $\xio \approx 0.7$, where the implications and shortcomings of fitting data with large age uncertainties are discussed extensively.
%%%%%%%%%%%%%%%%%%%%%%%%%%%%%%%%%%%%%%%%%%%%%%%%%%%%%%%%%
%
%
%
%          Section: Results
%
%
%
%%%%%%%%%%%%%%%%%%%%%%%%%%%%%%%%%%%%%%%%%%%%%%%%%%%%%%%

\begin{figure*}
\includegraphics[scale=0.6]{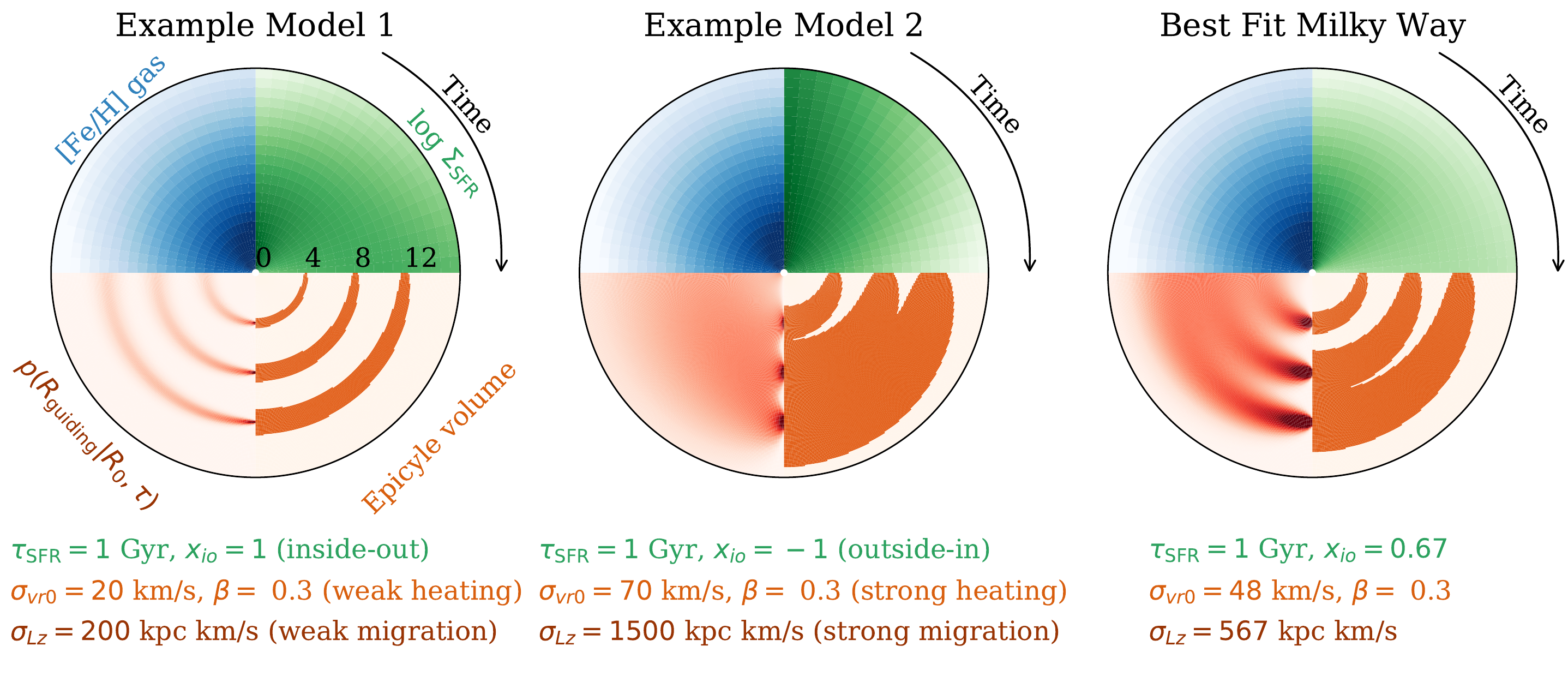}
\caption{Schematic illustration of the four main model aspects (left and middle panels), shown for the best-fit model in the right panel. Each panel represents a realization of the Milky Way disk model presented in Section~\ref{sec:model} with given model parameters (see legend). The radial axis in polar coordinates is Galactocentric radius. The angle in each quadrant represents time, increasing clockwise. For the top quadrants, time follows the evolution of the Galaxy. For the bottom quadrants, time is time since the birth of a star, tracing the evolution of its orbit. Each quadrant of the disk is color-coded by one of the model aspects. Going clockwise from top left, they display (i) the gradual enrichment of the gas in iron (top left, blue), (ii) the gradual build up of the disk (top right, green), (iii) the radial range occupied by stars born at 4, 8, and 12 kpc due to their radial motion (epicycle volume, bottom right, orange) and (iv) the probability density of the same stars to have their guiding radius at different places in the disk due to radial mixing (bottom left, red). At birth, stars are assumed to be on near circular orbits so, as can be seen in all panels, the three orbits born at 4, 8, and 12 kpc can be easily disentangled. Model 2 (middle) is undergoing strong radial mixing and strong radial heating, which mixes the orbits such that the {\it pdfs} of the stars cover almost the entire disk. Model 1 (left) has only modest radial heating and migration. The actual best fit model (right) falls in between these two regimes, where stars mix significantly, but not enough to erase all dynamical memory of their birth conditions.
%\com{HWR}{I love that figure! Is there any way, to enhance the (color) contrast in the top two quadrants? [I suspect you have tried; then never mind.}
 \label{fig:model_aspects}}
\end{figure*}

\section{Best fit Milky Way Disk Model \label{sec:results}}

Fig.~\ref{fig:results_corner} illustrates that all model parameters are well constrained, that the {\it pdfs} are approximately Gaussian, and that for most parameter combinations the covariances are small. The model seems to be well-posed for and well-constrained by the data set. This holds true for its dynamical aspects (brown in Fig.~\ref{fig:results_corner}), inside-out growth (green), and ISM enrichment (blue). The best fit model is illustrated in a schema in Fig.~\ref{fig:model_aspects}.

In the following subsections, we look at these different aspects more closely. 

\subsection{Migration strength and Age-Radial Velocity Dispersion}
The strength of radial migration is encompassed in the model aspect $\srm(\tau)$, the width of the distribution of stars of age $\tau$ about their mean angular momentum. We find $\srm(\tau) = 567~\mathrm{kpc~km/s} \sqrt{\tau / 6~\mathrm{Gyr}}$. With a circular velocity of about $235$ km/s, this corresponds to a migration scale of $2.4$ kpc for the 6 Gyr stars.

Radial heating leads stars to increase their random motion in the radial direction. Using Eq.~\ref{eq:heating},
we find that the velocity dispersion at the Sun is about 43 km/s for the $\sim$ 6 Gyr stars, in line with \cite{nordstrom_etal_2004} and \cite{mackereth_2019}, and that the age dependency is $\tau^{0.3}$. 
In the epicycle approximation, radial heating leads to epicycle amplitudes typically of $A_\mathrm{epi} = \sqrt{ \sigma _r^2(\tau)/\kappa ^2 }\approx 1.5 \mathrm{kpc}$ for the older stars, as illustrated in Fig.~\ref{fig:model_aspects}.
The parameter $R_{\sigma_R}$ best fit value is $\approx 27$ kpc, whereas it has been commonly assumed to be of order 2 $\times$ disk scale length ($\sim6-7$ kpc). The original motivation for the exponential decay of the radial velocity dispersion was to keep a disk with a constant scale height $h_z$ and a constant ratio $\sigma_r/\sigma_z$ \citep{van_der_kruit_1982}, with $\sigma_z \propto h_z\sqrt{\rho} \propto \exp(-R/2R_d)$. But (1) the agents driving the evolution of the vertical and radial motion are likely different \citep[e.g][]{sellwood2014} so $\sigma_r$ and $\sigma_z$ need not be related, and (2) the Galactic disk is now known and expected to flare \citep{ting_rix_2019, minchev_2015, bland-hawthorn_gerhard_2016, bovy_etal_2016, kawata_2017, sanders_das_2018, mackereth_etal_2017, mackereth_2019}: the scale-height of coeval stellar populations increases with radius, hence there is no need for $R_{\sigma_R}$ to be small.

\subsection{Inside-out Growth and the Metallicity Profile}

The model fit favors a global inside-out growth of the disk, where stars formed first from low angular momentum gas, and star formation moved gradually to higher angular momentum ($x = 0.68$). This is in accord with the results of \cite{frankel_etal_2019}, who used a similar model in radius on the APOGEE-RC data set of the 12th data release. Our model is illustrated for different values of $x$, as well as for the best fit, in Fig.~\ref{fig:model_aspects} in green (top right quadrants). As the disk forms from inside-out, the enrichment in metals (here iron, [Fe/H]) proceeds with a radial gradient and still on-going enrichment (Figure \ref{fig:tagging_birth_feh}).

\subsection{The Orbit-Age-Abundance Distributions}

The distributions most directly affected by radial mixing are the metallicity distribution functions of stars at given radius $p(\fe |R)$ \citep{hayden_2015, loebman_etal_2016}, whose shape is influenced by the amount of metal-rich stars incoming from the inner disk the metal-poorer stars coming from the outer disk, and the initial metallicity profile of the gas from which stars formed, set by the inside-out star formation history \citep{schoenrich_mcmillan_2017, schonrich_binney_2009a}.

\begin{figure*}
    \centering
    \includegraphics[width=\textwidth]{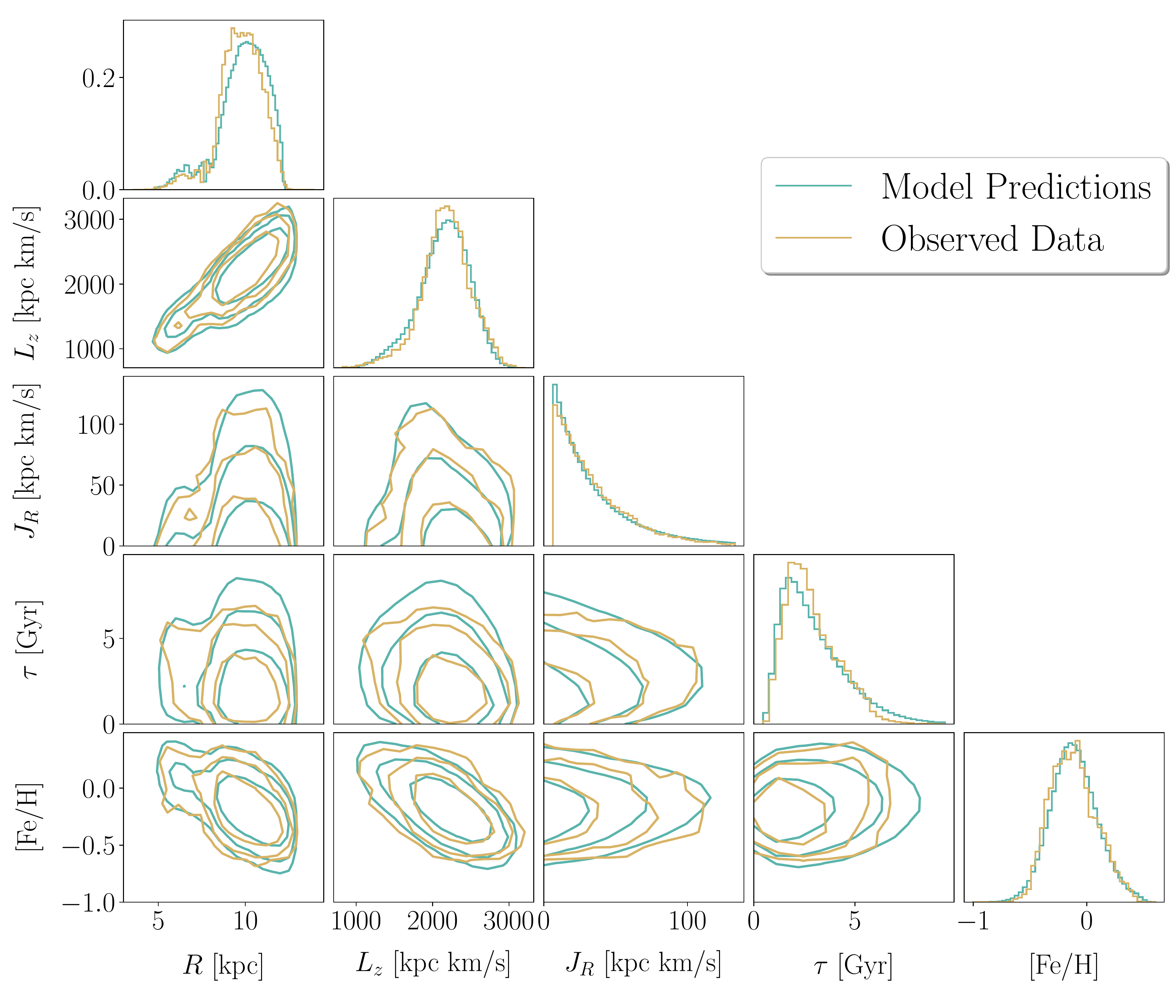}
    \caption{Best fit model predictions in data space (brown), generated from the MLE estimates of Table \ref{table:model_aspects}, compared with observed densities in APOGEE DR14 data (brown). The model agrees remarkably well with the data: the 1D distributions are generally well reproduced, but more importantly the 2D dististributions and the entire shape of the density the 5 dimensional data space agrees well. The age-metallicity-orbit structure of the dataset seems well recovered.}
    \label{fig:data_space_comparisons}
\end{figure*}

The available data, and the framework we developed in this work, allow us to make comparisons in more dimensions, reproducing the entire data set's 5-dimensional distribution $p(\fe, \tau, J_R, L_z, R)$ resulting from the global evolution of the Milky Way disk (and selection procedures).
Figure \ref{fig:data_space_comparisons} illustrates the distributions $p(\fe, \tau, J_R, L_z, R)$ observed in the data set (brown) and predicted by the best fit model for the data set (green). Overall, the observed distributions are globally well recovered by the model fit. The metallicity radial profile (bottom left panel of the corner plot in Fig \ref{fig:data_space_comparisons}), which arises from (1) chemical evolution and (2) subsequent orbit evolution is well described, and illustrates that the metallicity distribution functions $p(\fe, R)$ and the age-metallicity distributions (4th panel, bottom), usually the main diagnostic used in the literature, are all well reproduced. The distributions in $L_z$ and $J_R$ reflect both the secular evolution of the disk and APOGEE's spatial selection function.

%
%
% Subsection tests
%
%
\subsection{Verifications: Model Variants and Parameter Recovery \label{subsection:tests}}
During the construction of this model, we tested a set of model variants. We first tested the parameter recovery on noised mock data, and rejected all models for which we could not recover the true parameters. For example, we could not add a parameter to quantify how much angular momentum is actually conserved while stars diffuse in angular momentum (by e.g., fitting for a simple form of diffusion coefficient $D^{(1)}$ from Eq.~\eqref{eq:diffusion}) because there is a degeneracy with the enrichment model.

Once the best model candidate (which is presented in Section \ref{sec:model}) was identified, the best fit parameters found with MLE, and the posterior sampled with MCMC, we performed a series of additional tests to verify different aspects of our results. We investigated whether uncertainties in our integral calculations introduced systematics or biases in the estimation of the parameters, by computing it using different model realizations (by changing the model parameters) and Monte Carlo samples of different sizes, and found no change in the results within 1$\sigma$ (for both the data used in the present paper and mock samples generated from our model). We generated and fitted mock data with different noise levels (increasing or decreasing the formal uncertainties by a few percent), and the best fit parameters are well recovered too.

We studied the effect of varying the model of the potential on the estimates of the dynamical parameters. Using the default {\sl MWPotential2014} in {\it Galpy} with a circular velocity of 220 km/s at 8 kpc, we found small changes in the estimates of \srm. These changes are however expected and quantifiable.  The estimate for radial migration dropped by about $\sigma_{Lz220} \approx \srm \times \frac{220}{235}$, which is what one expect since the metallicity scatter, and the radial metallicity gradient in the stars are the same, and $\srm \approx \sigma_\fe v_\mathrm{circ}/{\dfe}$. 
But a full exploration of alternate gravitational potentials is beyond the scope of the present work. The potential model we are currently using is well constrained by external data \citep{bovy_rix_2013, bovy_2015_galpy}.

\subsection{Model Limitations} \label{subsection:model_limitations}

Forward modeling the orbit-age-[Fe/H] structure of the APOGEE$\times$Gaia dataset, with an interpretable model that accounts simultaneously for diverse aspects of Galaxy evolution, data uncertainties and the survey selection function, provides a framework with a great potential for Galactic archaeology. The best fit model is well tested and reproduces the observed trends and distributions of the data set well. However, at present our model lacks some features that one might desire from a full physical model of the Galaxy:

\begin{enumerate}
    \item Radial migration of stars in the Milky Way is measured indirectly through its impact on the age-metallicity distributions. The strongest assumption we have made in that direction is that the birth age-metallicity relation was tight and monotonic, and that our choice of functional forms to model the evolution of [Fe/H] were flexible enough and adequate. However, any inadequacy or inappropriate rigidity in that model will be measured as radial migration in this context, so $\srm$ could be a lower limit on the strength of radial migration.

\item Our description of $L_z$ change is only an approximation of the solution to the diffusion equation (Eq.~\ref{eq:diffusion}) that is valid far from the Galactic center, and is not self-consistent. It is likely that the strength of migration is a function of radius and time  \citep[e.g.][]{kubryk_etal_2013, toyouchi_chiba_2018}, which in our simple diffusion picture is ignored. Additionally, Eq.~\ref{eq:diffusion} should contain a source term for star formation, which we have modeled separately. This should not impact the results drastically, as the two extreme regimes are recovered: in the limit where radial migration is asymptotically inefficient, stars' current $L_z$ distribution is a Dirac function of their birth distribution, and the overall $L_z$ distribution is the exponential profile used for initialization in Eq.~\ref{eq:Lz0_distribution}. In the limit where radial migration is asymptotically efficient, our imposed steady state solution for the $L_z$ distribution recovers the results of \cite{herpich_etal_2017} for a flat circular velocity curve, with the same scale length \rdo.

\item The secular evolution processes in the disk (diffusion in $L_z$ and increase in $J_R$) were treated independently. However, they should be covariant depending on the heating agents, (e.g. as in Eq.~\ref{eq:sellwood_binney_delta_Jr}) \update{and because a star's chances to be trapped and corotation resonance, and thereby migrate radially, should depend on its radial and vertical motions \citep{daniel_2019, solway_2012}}. Ideally, we should treat the entire Fokker-Planck equation in action space. However, we argue that this should have only a small impact on the present results, because the disk remains relatively cold ($J_R$ does not increase much) and $L_z$ diffusion is stronger by an order of magnitude \update{and in this modelling, both heating and migration are conditioned on time, making their conditioning on each other only implicit}. This may blur out such covariances on large spatial and time-scales. However, in the Solar neighbourhood, there are indeed clear over-densities in the $L_z-J_R$ plane, some of them arising along Lindblad resonances with the bar or other non-axisymmetries \citep{sellwood_binney_2002, trick_coronado_rix_2019,  trick_2019}). These portray the impact of the most recent set of non-axisymmetries.

\item The treatment of radial heating, with the time dependence $\beta$ could depend on angular momentum. Indeed, as shown in simulations, the heating time dependence $\beta$ could depend on spiral arms and Galactocentric radius \citep{aumer_2016, BT08}. The radial dependence of $\beta$ in the age-velocity dispersion relation in the Milky Way disk was confirmed in \cite{mackereth_2019}. The present model fit leads to $\beta \approx 0.3$, which is typical and agrees with the largest part of the disk values in \cite{mackereth_2019}, but not in the outer disk where their $\beta $ decreases, possibly due to the weakening of spiral strength.

\end{enumerate}

%%%%%%%%%%%%%%%%%%%%%%%%%%%%%%%%%%%%%%%%%%%%%%%%%%%%%%%%%
%
%
%
%          Section: Discussion and conclusions
%
%
%
%%%%%%%%%%%%%%%%%%%%%%%%%%%%%%%%%%%%%%%%%%%%%%%%%%%%%%%%%
\begin{figure}
    \centering
    \includegraphics[width=\columnwidth]{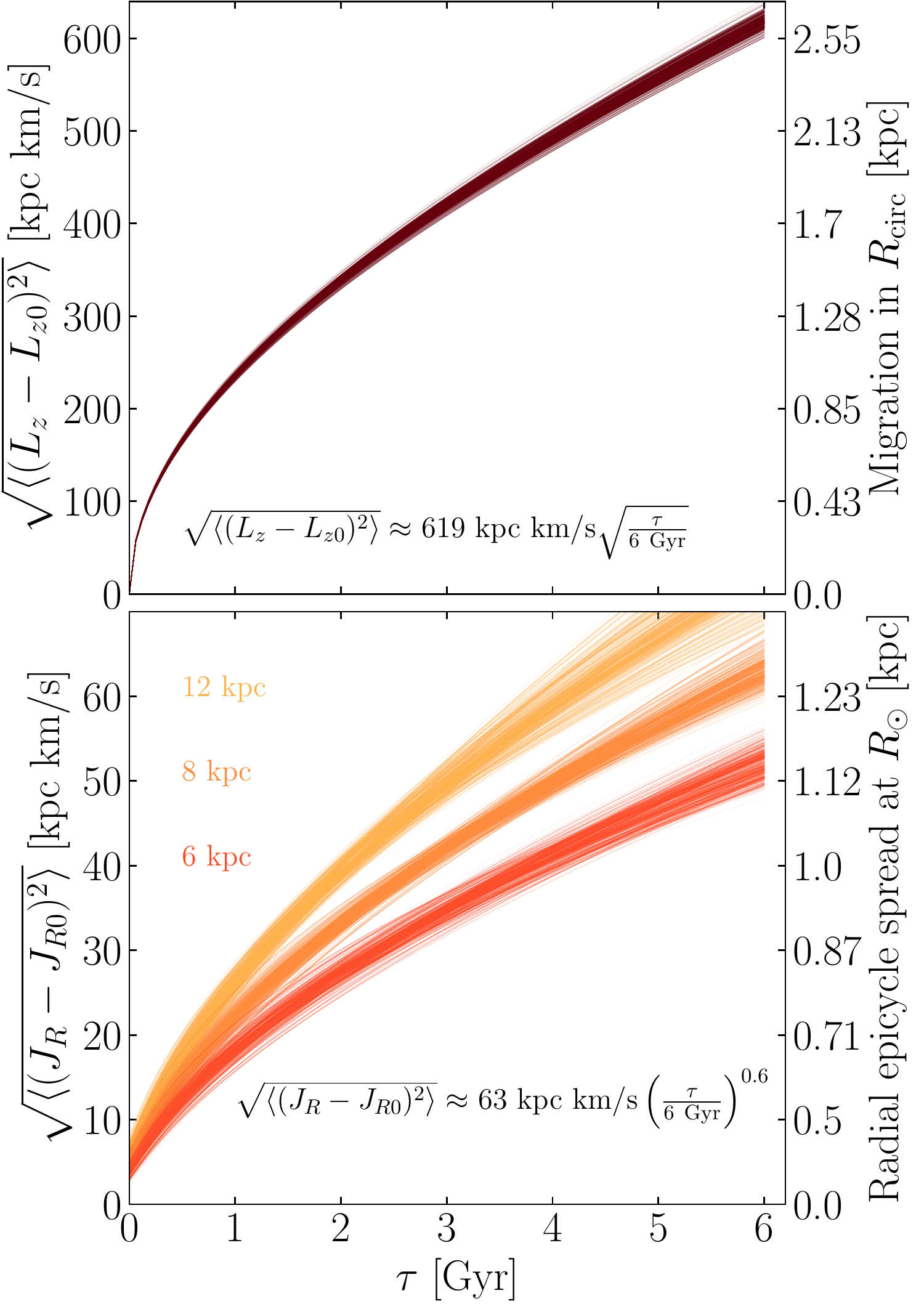}
    \caption{Secular evolution of the Milky Way disk implied by the best fit model, as diffusion in angular momentum (top) and radial action (bottom) drawn from the MCMC samples illustrated in Fig \ref{fig:results_corner} and using the model equations, Galactocentric radii 6, 8 and 12 kpc. The right-hand side y axis shows the equivalent in distance units. Top: we assumed a constant circular velocity curve $R_\mathrm{circ} = L_z/v_\mathrm{circ}$ with $v_\mathrm{circ} = 235$ km/s. Bottom, shown for the Solar Galactocentric radius: we assume the radial spread about the guiding radius due to the epicycle motion corresponding to a radial action $J_R$ at $R$ is $\sqrt{\langle (R-R_g)^2 \rangle} = \sqrt{J_R/\kappa} = A/\sqrt{2}$ with $\kappa$ the epicycle frequency and $A$ the epicycle amplitude.}
    \label{fig:secular_evolution}
\end{figure}
\section{Astrophysical Implications and Discussion \label{sec:discussion}}

% Secular processes are important, have significant effects, but are not well quantified in the Milky Way disk 
Secular processes drive stellar orbits to evolve, which can have a large impact on a disk galaxy's evolution. If strong, such diffusion processes lead to a near complete dynamical memory loss, challenging efforts in Galactic archaeology to try to infer a galaxy's history from its present-day properties. Radial migration is such a process and was shown to be strong in simulations \citep[e.g.][]{minchev_etal_2013}, but had not been well quantified across the Milky Way disk.

% Our model is awesome and dataset (almost) up to date
We have presented the first global model relating stellar ages, chemistry, and dynamics, which we have fitted to high-quality data over a large extent of the Galactic's low-$\alpha$ disk, accounting rigorously for data uncertainties and selection effects. The model builds on and extends \cite{sanders_binney_2015}'s pioneering framework. The previous modeling to \cite{frankel_etal_2018} was based on more local data and did not incorporate stellar ages. In this work, we have forward-modelled the distributions of stellar ages, [Fe/H], $L_z$ and $J_R$ with a radially-dependent star formation history accounting for the inside-out growth of the disk, a parametric chemical evolution model, and birth and action-based radial heating and radial migration orbit-evolution model. We have fitted all aspects of this model simultaneously, but focus on the orbit evolution, and treated the other model aspects as a nuisance. We have used the APOGEE$\times$Gaia red clump sample, a dataset that covers a large part of the Galactic disk, from $R \sim 4$ kpc to 13 kpc, a scale larger than the typical radial migration scale of a few kpc.

Large efforts have been made to measure the strength of radial migration in the Milky Way, but previous works lacked data on large spatial scales, making quantifying a large scale diffusive process a hard task. Furthermore, previous approaches did not use stellar ages \citep[e.g.][]{sanders_binney_2015} instead relying more tangentially on the known local relationships between stellar age and kinematics. Other methods measured a scatter in the [Fe/H]-[$\alpha$/Fe]-$R$ directly, but without accounting simultaneously for the galactic evolution processes that could contribute to it, thus lacking a framework that could use and describe the entire dataset \citep[e.g.][]{hayden_2015}. In \cite{frankel_etal_2018, frankel_etal_2019}, we developed a framework accounting for the main evolution aspects of the Milky Way disk, with a radially-dependent star formation history, chemical evolution of the disk, and evolution of the stars' Galactocentric radius. This model constrained a global orbit migration scale of about 3 kpc $\sqrt{\tau/6~\mathrm{Gyr}}$, implying that radial mixing happens on scales comparable to the scale-length of the Milky Way disk. However, this description of the disk only measured diffusion in Galactocentric radius, and not angular momentum, so failed to disentangle the two major processes causing stars to change Galactocentric radius (`churning' and `blurring').

\subsection{Secular Dynamical Evolution}
Our model describes the in-plane secular evolution of the Milky Way's low-$\alpha$ disk and disentangles the contributing processes: diffusion in angular momentum (`churning') and increase in radial action (`blurring', or `radial heating'). As both processes are diffusion in action space, we can quantitatively compare their strengths in a meaningful way, and here we choose to inspect the root-mean-squared (rms) deviation in the actions.

We first work out the rms deviations of the actions expected in the Solar neighbourhood from external data, and will then show that our more global model recovers this particular case. From a simplistic perspective, the rms deviation in the radial action is related to the disc properties as
\begin{equation}
    \sqrt{\langle(\Delta J_R)^2\rangle}\approx\frac{\sqrt{2}\sigma_R^2}{\kappa}.
\end{equation}
Taking the `textbook' quantities for the radial epicyclic frequency $\kappa\approx37\,\mathrm{km/s/kpc}$ and $\sigma_R\approx38\,\mathrm{km/s}$ for old stars from \cite{BT08}, we find $\sqrt{\langle(\Delta J_R)^2\rangle}\approx55\,\mathrm{kpc\,km/s}$. Likewise, the rms deviation in the angular momentum can be simply expressed as
\begin{equation}
    \sqrt{\langle(\Delta L_z)^2\rangle}\approx\frac{v_\mathrm{circ}\sigma_\mathrm{[Fe/H]}}{|\mathrm{d[Fe/H]}/\mathrm{d}R|}.
\label{eqn::lzsimple}
\end{equation}
Using approximate values for the solar neighbourhood of $\sigma_\mathrm{[Fe/H]}=0.2\,\mathrm{dex}$ for the metallicity dispersion \citep{nordstrom_etal_2004}, $|\mathrm{d[Fe/H]}/\mathrm{d}R|=0.062\,\mathrm{dex/kpc}$ for the radial metallicity of young stars \citep{lucklambert2011} and $v_\mathrm{circ}=235\,\mathrm{km/s}$ we find $\sqrt{\langle(\Delta L_z)^2\rangle}\approx750\,\mathrm{kpc\,km/s}$. These simple calculations confirm that plausibly $\sqrt{\langle(\Delta L_z)^2\rangle}$ is an order of magnitude larger than $\sqrt{\langle(\Delta J_R)^2\rangle}$.

Expanding to the greater extent of the disk with our full model, from Eq.~\ref{eq:p_radial_migration}, the variance of the angular momentum distribution of a stellar population of age $\tau$ is
% about the Galactic center as a function of time, resulting from $L_z$ mixing processes is
\begin{equation}
   \langle (L_z - L_{z0})^2 \rangle = \sigma(\tau)^2 + (D\tau)^2.
\end{equation}
where 
the drift term $(D \tau)^2$ is subdominant, contributing only  $\sim20$\% to $\langle (L_z - L_{z0})^2 \rangle $.

Similarly, the variance of the radial action of a population of age $\tau$ is
\begin{equation}
\langle (J_R - J_{R0})^2 \rangle = 2 \langle J_R \rangle^2 + 2J_{R0}^2 -2 J_{R0} \langle{J_R}\rangle,
\end{equation}
with $J_{R0}$ the radial action at birth, which we assume here is zero since in the model stars are born on near-circular orbits. $\langle J_R \rangle = \sigma_R^2/\kappa$ is the mean radial action as defined in Eq.~\ref{eq:p_radial_action}. 
Both of these quantities are plotted as functions of $\tau$ in Fig.~\ref{fig:secular_evolution}. Using a reference age of $\tau_m=6\,\mathrm{Gyr}$, we find at 8 kpc (see Fig. \ref{fig:secular_evolution} for the spatial variations)
\begin{equation}
\begin{split}
\sqrt{\langle (L_z - L_{z0})^2 \rangle} &\approx (619~\mathrm{kpc~km/s})
\left( \frac{\tau}{\mathrm{6~Gyr}}\right) ^{0.5}
,\\
\sqrt{\langle (J_R - J_{R0})^2 \rangle} &\approx (63~\mathrm{kpc~km/s})
\left( \frac{\tau}{\mathrm{6~Gyr}}\right) ^{0.6}.
\end{split}
\end{equation}
We note that up to a factor between 1.2 and $\sqrt{2}$, these quantities are very close to \srm=572 kpc km/s ~and the mean radial action $\langle J_R \rangle=\sigma_R^2/\kappa = 45$kpc km/s respectively, so our general conclusions do not depend much on the details of our choice of reference quantities (i.e. \srm versus $\sqrt{(\Delta L_z)^2}$).
% }

A spatial representation of the diffusion in angular momentum and increase of radial action is illustrated in Fig.~\ref{fig:model_aspects}. We show two examples for the secular evolution of the disk for the first two panels (weak mixing and heating, and strong mixing and heating), and the best fit in the third panel. The second $y$ axis of Fig.~\ref{fig:secular_evolution} also illustrates this more quantitatively.

Using the same APOGEE RC dataset, \cite{frankel_etal_2018, frankel_etal_2019} measured a migration strength in Galactocentric radius of $3.1~ \mathrm{kpc} \sqrt{\tau / \mathrm{6~Gyr}}$, which is slightly larger than we have found here. These models were purely spatial and ignored the dynamics. The spreads in the metallicity distributions in these models are wholly accounted for by the radius migration and its strength is more simply linked to the radial metallicity gradient (as in Eq.~\eqref{eqn::lzsimple}). In the new dynamical model presented here, spreads in the metallicity distribution are due to a combination of both migration and heating, the latter of which introduces more extreme metallicity stars from the inner and outer disc at their apo- and pericentres respectively, and so further broadens the metallicity distributions. This reduces our measured radial migration strength to $\sqrt{\langle(\Delta L_z)^2\rangle}/v_\mathrm{circ}=2.6~ \mathrm{kpc} \sqrt{\tau / \mathrm{6~Gyr}}$ with the difference coming from the radial heating.

%
% Subsection : Implications of a strong Lz diffusion process
%
\subsection{Implications of a Strong $L_z$ Diffusion Process}

If the strength of angular momentum diffusion, of the order of the mean angular momentum of the Galaxy, is typical to all disk galaxies, this redistribution has important implications for galactic archaeology for external galaxies. For the Milky Way, the strength of radial migration can be measured through a physical [Fe/H]-age-$L_z$ scatter, obtained by data for individual stars. Such a framework may not be applicable in external galaxies beyond the Local Group, where all properties are integrated. Studying stellar populations in external galaxies may lead to good present-day age histograms and present-day mass-weighted age gradients, but reflect only mildly the formation of galaxies due to important dynamical memory loss: i.e. age and metallicity radial gradients weaken \citep{frankel_etal_2019}.

As argued in \cite{herpich_etal_2017}, an asymptotically strong redistribution of stellar angular momenta in cold disks could naturally lead disk profiles to follow exponential distributions, as is observed in disk galaxies \citep[e.g.,][]{deVaucouleur_1948, freeman_1970ApJ}. Since disks are not always expected to form with exponential profiles \citep[e.g.][]{roskar_etal_2008a}, a strong diffusive process that leads to an exponential profile irrespective of the initial conditions
could reconcile the observations with simulations of galaxy formation.

Even though mixing processes are strong, they are not strong enough to erase all gradients in which case even using chemical-age information would not rewind stars back to their birth conditions because the final state of the system would be independent from its initial state. In the Milky Way, metallicity and age radial gradients are weakened, but not erased.

However, any modelling of the Milky Way's chemical evolution requires us to account for the strong radial orbit redistribution. For instance, a local age distribution might reflect better the global star formation history of the disk rather than the local star formation history, as a local sample of stars, even those on circular orbits, may contain stars born kiloparsecs away and lack stars born locally. 

\subsection{Disentangling $L_z$-Diffusion from Heating}

Dynamical processes produce correlated changes in the actions of stars. Assuming $\Delta J_R = f(\Delta L_z)$, the general function $f$ will depend on the specifics of the dynamical interaction.
In the present work, we do not explicitly model the possible interactions leading to radial heating and angular momentum diffusion; we only measure their effect over 6 Gyr of evolution with an effective model, and find that across the disc
\begin{equation}
\sqrt{\langle (J_R - J_{R0})^2 \rangle} \approx 0.1 \sqrt{\langle (L_z - L_{z0})^2 \rangle}.
\end{equation}
This result can already provide some global constraints on the nature of dynamical processes across the Milky Way, but not on the details of the secular interactions. The net changes are not directly comparable to the changes expected over single migration events $\Delta J_R = f(\Delta L_z)$.
To zeroth order, near the main resonances of a non-axisymmetry rotating at a constant pattern speed $\Omega_P$ (e.g. the bar or a spiral wave), 
the change of radial action is related to the change in angular momentum through \citep[e.g.][]{sellwood_binney_2002}
\begin{equation} \label{eq:sellwood_binney_delta_Jr}
 \Delta J_R = \frac{\Omega - \Omega_P}{\kappa} \Delta L_z.   
\end{equation}
Here $\kappa$ and $\Omega$ are the radial and azimuthal frequencies. \cite{sellwood_binney_2002} point out that near corotation ($\Omega = \Omega_P$), $\Delta J_R$ should be very small even though $\Delta L_z$ can be large -- a star can move from circular orbit to circular orbit. Therefore, there is no \emph{dynamical} evidence that a star found on a near circular orbit at radius $R$ today was not born on a different circular orbit. Around the Lindblad resonances, where $\kappa = \pm m(\Omega - \Omega_P)$, interactions with non-axisymmetries tend to heat the disk $\Delta J_R=\pm\Delta L_z/m$.
More recent works argue that $\Delta J_R = f(\Delta L_z)$ is not necessarily linear, and that angular momentum redistribution at corotation might not always occur without changes in $J_R$: resonances can overlap, leading to non-linear effects and stochastic motions of the stars \citep[e.g][]{minchev_2011, minchev_2012, daniel_2019}. In this model context, our findings suggest that migration near corotation was important.

In addition to these non-linear effects, $\Delta L_z$ and $\Delta J_R$ may deviate from Eq.~\ref{eq:sellwood_binney_delta_Jr} if 
 spiral perturbations do not rotate as solid bodies as seen in simulations \citep[i.e. with a pattern speed that changes with Galactocentric radius][]{quillen_2011, grand_etal_2012} or indirectly with extragalactic observations \citep{merrifield_2006,masters_2019}, or if their pattern speed is a function of time.

The simulations described above, as well as those of \cite{brunetti_etal_2011} and \cite{loebman_etal_2016}, have brought understanding of the processes involved in the secular evolution of disk galaxies and the processes at play in radial orbit migration, and have pioneered qualitative comparisons with Milky Way data. However, they are not directly comparable to observed data in the Milky Way because observed data are noisy and do not represent the full Milky Way disk. More importantly, the present view of the Milky Way only represents the equivalent of the final snapshot of a simulation, as argued in \cite{aumer_2016}. This means age-kinematic relations differ from heating histories (as much as time differs from age) and our effective models cannot be used to recover robustly the full evolutionary history of the Galaxy. Simulations are necessary to guide the construction of realistic and physically motivated forward models and to make the link between the global measure of $\sqrt{\langle (J_R - J_{R0})^2 \rangle}$, $\sqrt{\langle (L_z - L_{z0})^2 \rangle}$ and the instantaneous changes $\Delta J_R$ and $\Delta L_z$.

\subsection{Implications for the Sun and the Solar System} 

In a model with significant radial migration, the Sun potentially formed quite far from its present Galactocentric radius. Here we analyse the most likely history of the Sun using our model.

Using Eq.~\ref{eq:eq_Fe_Ro_t} with the Sun's age of $\tau_\odot = 4.6$ Gyr \citep{bonanno_2002_agesun} and $\fe_0 = 0\pm 0.05$ dex \citep{asplund_2009}, 
we find the birth angular momentum of the Sun was $L_{z0\odot} \approx 1824 \pm 127$ kpc km/s . If, at the time of its formation, the circular velocity corresponding to this angular momentum were 235 kpc km/s, this would correspond to a birth Galactocentric radius of $7.8 \pm 0.6$ kpc, which is 5\% closer to the Galactic center than today. This is quite different from previous estimates of the Solar birth location in \cite{frankel_etal_2018} (5.3 kpc), but in better agreement with \cite{minchev_2018} (7.3 kpc) and \cite{haywood_etal_2019}. We interpret this significant change of Solar birth location from our previous estimates as a consequence of two model modifications. Firstly, the introduction of the drift term $D$ towards the inner disk in the diffusion equation \ref{eq:diffusion}. In \cite{frankel_etal_2018}, $D$ was set to zero, which resulted in global outwards migration due to the negative density gradient, with a disk profile that broadens with time. Here, we approximately conserve angular momentum, with a disk profile remaining approximately constant over time. As a result, stars have a higher probability to migrate inwards than outwards. Secondly, our chemical enrichment description (Section~\ref{section_enrichment}) is different: it is a function of birth angular momentum and not birth Galactocentric radius, and we have imposed a flattening of the [Fe/H] profile in the inner disk, which is more physically and observationally motivated.

The Solar birth Galactocentric radius is still widely debated and not well constrained. Most chemical evolution arguments lead to birth radii estimates between its present-day radius and 3 kpc closer to the Galactic center \citep[e.g.][]{wielen_1977, nieva_przybilla_2012, minchev_2018, sanders_binney_2015, feltzing_2019, kubryk_etal_2015a, frankel_etal_2018}, except for \cite{haywood_etal_2019} who argue that the Sun is a typical outer disk star. Even though different models infer different birth radii for individual stars, the overall radial migration rate estimate remains similar in all models, as shown in \cite{feltzing_2019}. Finally, \cite{martinez-barbosa_etal_2015} use backward integration over the Sun's lifetime, concluding that the Sun was born in the outer disk. However, without knowledge of the past evolution of the Milky Way's potential, such an exercise is not trivial.
Better estimates of the Solar birth place may additionally allow to put tighter constraints on the environment in which the solar system has evolved \citep[e.g. encounters with Giant Molecular clouds][]{kokaia_davies_2019} \update{even though knowing both the Solar birth  and current orbits does not imply that the Sun has always remained between the two: it could well have migrated back and forth to the same place, since the typical migration distance for a 4.6 Gyr old star is about 2 kpc}.

% Implications for the distributions of stars born with the Sun
\subsection{Application to the Solar Siblings' Orbit Distributions}
%The Sun formed together with a cohort of solar siblings Chemical identification of these siblings allows for accurate characterization of the dynamical disruption of the Sun's birth cluster, and its subsequent dispersal throughout the Galactic disc. However, it is not trivial to identify the siblings. In our model, we have, in essence, performed this task in an averaged sense for many stars across the disc, enabling us to measure the average dynamical history of a population born at a given location and time. We can now use our model to find the most likely history of the solar siblings.

We estimate the possible present-day $L_z$ and $J_R$ ranges occupied by stars that were born with the same $\Lzo$, the same \fe, and at the same time as the Sun with Eq.~\ref{eq:p_radial_migration}. Assuming solar siblings undergo phase mixing rapidly (the Sun has undergone $\sim$ 20 Galactic orbits), there is then no dynamically noticeable difference between `sharing the same birth cluster as the Sun' and `being born with the same \Lzo, and time'. From our model fit, 95\% of these stars should currently have $550 \leq L_z \leq 2770$ kpc km/s and $J_R \leq 130$ kpc km/s. This is roughly consistent with the results of \cite{webb_etal_2019}, who used simulations to investigate the present-day positions of solar siblings in the $(L_z, J_R, J_z)$ space in different possible potentials and constrained present-day solar siblings angular momenta to $353 \leq L_z \leq 2110$ kpc km/s and $J_R \leq 116$ kpc km/s. The exact values of these bounds should depend on the detailed history of the Milky Way disk, but their model gives an angular momentum range of about 2000 kpc km/s, which is close to our 2$\sigma(\tau)$ value.

However, if the abundance profile of the gas in the Galactic disk is really axisymmetric, then there is no chemical information on the phase to disentangle whether a star is born in the same birth cluster as the Sun, or just at the same Galactocentric radius (or \Lzo). \update{Therefore, the Sun could well be born from a birth cluster (possibly now disrupted) that is different from the candidate cluster M67, which has similar age and metallicity as the Sun \citep{yadav_2008,heiter_2014}: it was shown unlikely to be the Sun's birth cluster, but not fully ruled out \citep{jorgensen_church_2020, webb_etal_2019}}. The recent analysis of \cite{ness_etal_2019} shows that stellar orbits and abundances can be well predicted with only [Fe/H] and age, implying that our present analysis contains the most essential elements for chemical tagging.

\subsection{Limitations and caveats}

The physical limitations of our modeling were discussed extensively while presenting the best fit Milky Way disk model in Subsection~\ref{subsection:model_limitations}: the model could improve by allowing a time- and radius-dependent strength of radial migration and radial heating.

We now discuss another approximation we have made while constructing the model: the model for the
population selection of the red clump stars. In practice, the red clump selection is based on a neural network trained to classify stellar evolutionary stage from their spectra, trained on asteroseismic data \citep{ting_hawkins_rix_2018}. In the model, we have approximated the selection of red clump stars as a cut in logg-Teff-color space as in \cite{bovy_etal_2014}. This approximation is well motivated because (1) the classifier is currently one of the best methods to obtain a pure and complete sample of red clump stars given their spectra \citep{ting_hawkins_rix_2018}, and (2) the modeled cut in logg-Teff-color space is, in theory, a good approximation to selecting core helium burning stars.

However, this approximation is conceptually not satisfying (as for other methods based on data driven selection functions), and with no known applicable and rigorous solution: our treatment of the population selection function in the model is conceptually inconsistent with the actual selection of stars.  
The actual data-driven selection of stars is not trivial to forward model: this method takes a star's spectrum as input and returns asteroseismic parameters $\Delta \nu$ and $\Delta P$, thereby disentangling RC and secondary RC. Therefore, to assess selection effects in our forward model properly, we should generate a set of theoretical spectra of various ages, $T_\mathrm{eff}$, $\log g$ and various abundances, add noise and instrumental effects, and then pass these spectra to the neural networks that selected the red clump stars to evaluate the fraction of generated stars  that becomes classified as red clump, as a function of age and metallicity (or any stellar quantity that we wish to model). Hence, if for example the neural networks were to fail in some areas of the parameter space (where e.g. the training data are sparse), the effects would be fully modelled.

As machine learning-based selection functions are (justifiably) becoming important in the field of astrophysics, they will also need come with a consistent modeling framework in the future.

Instead, we have approximated this population selection effect as a function of age from \cite{bovy_etal_2014} as in \cite{frankel_etal_2018, frankel_etal_2019} and assumed it was sufficient for the purpose of our work. We argue qualitatively that this should have only a small impact on the present work: such a selection should mainly affect the distribution of ages (at large ages, were the training sample was sparse) of our sample, that is most closely linked to the star formation history in our modeling context. However, we treat this star formation history as a nuisance aspect and marginalize over it, and assume the red clump selection does not bias significantly the dynamical parameters of the stars (beyond the covariances between age and kinematics, which we do model), and we focus on radial migration and diffusion in action space. As can be seen in Fig.~\ref{fig:results_corner}, the dynamical parameters are not correlated with the star formation history.

%%%%%%%%%%%%%%%%%%%%%%%%%%%%%%%%%%%%%%%%%%%%%%%%%%%%%%%%%
%
%
%
%                Section: Summary
%
%
%
%%%%%%%%%%%%%%%%%%%%%%%%%%%%%%%%%%%%%%%%%%%%%%%%%%%%%%%%%
\section{Summary}
We have presented and applied a global model for the secular evolution of the Milky Way's low-$\alpha$ disk. The model accounts for an inside-out star formation history, [Fe/H] gradual enrichment, and the subsequent evolution of a stars' orbits as diffusion in action space. Applying this model to the APOGEE red clump stars, we have fully accounted for the selection function of the survey and for data uncertainties. The data are constraining, and the model fit with MCMC implies, in this context, that
\begin{enumerate}
\item $L_z$ redistribution evolves as  $\sqrt{\langle (L_z - L_{z0})^2 \rangle} \approx (619~\mathrm{kpc~km/s})\left(\frac{\tau}{\mathrm{6~Gyr}}\right)^{0.5}$, which corresponds to a migration distance of about 2.6 kpc for the 6 Gyr old stars (see Fig.~\ref{fig:secular_evolution}). In other words, for a coeval population of stars, 68\% of them will be within $ 2.6~\mathrm{kpc}\sqrt{\tau{/\mathrm{6~Gyr}}}$ of their birth radius and the remaining 32\% will have migrated further;
\item $J_R$ evolves as $\sqrt{\langle (J_R - J_{R0})^2 \rangle} \approx$\\ $ (63~\mathrm{kpc~km/s})
\left( \frac{\tau}{\mathrm{6~Gyr}}\right) ^{0.6}$. This shows that redistribution in angular momentum is stronger than increase of radial action, by a fator $\sim 10$, leading us to conclude that radial migration dominates the evolution of the Galaxy's low-$\alpha$ disk. This leads the disk to remain kinematically cold, but with a strong dynamical memory loss, making it necessary to use chemical and age information to recover the birth conditions;
\item the Sun's birth angular momentum inferred from the best fit is relatively close to its present-day angular momentum with $L_{z0,\odot} \approx 1824 \pm 127$ kpc km$s^{-1}$, but its siblings may have a large distribution in action space with a width of 2000 kpc km/s in $L_z$ and 130 kpc km/s in $J_R$.
\end{enumerate}
We have demonstrated that our approach can disentangle the diversity of dynamical phenomena that have shaped the Milky Way's disk. However, our parametric model is purely effective so does not capture the real complexities of individual dynamical processes, but only their average effect on the Milky Way.
We hope that this may be solved in the future, by applying and coupling this model to more detailed simulations of galaxy evolution.

\section*{Acknowledgements}
It is a pleasure to thank Michele and Georges Laillet for hosting the writing of early drafts of this work. 
We thank Paul McMillan, James Binney, Gregory Green, Paola Di Matteo and Misha Haywood for interesting discussions.

N.F. acknowledges support from the International Max Planck Research School for Astronomy and Cosmic Physics at the University of Heidelberg (IMPRS-HD). 
J.L.S acknowledges the support of the Royal Society, and the Leverhulme and Newton Trusts. 
H.-W.R. received support from the European Research Council under the European Union's Seventh Framework Programme (FP 7) ERC Grant Agreement n. [321035]. YST is grateful to be supported by the NASA Hubble Fellowship grant HST-HF2-51425 awarded by the Space Telescope Science Institute.

The following softwares were used during this research: Astropy \citep{astropy}, Matplotlib \citep{matplotlib}, {\sl Galpy} \citep{bovy_2015_galpy}, Emcee \citep{foreman-mackey_2013}.
Figure \ref{fig:results_corner} was produced using the package Corner \citep{corner}.

This work presents results from the European Space Agency (ESA) space mission Gaia. Gaia data are being processed by the Gaia Data Processing and Analysis Consortium (DPAC). Funding for the DPAC is provided by national institutions, in particular the institutions participating in the Gaia MultiLateral Agreement (MLA). The Gaia mission website is \href{https://www.cosmos.esa.int/gaia}{https://www.cosmos.esa.int/gaia}. The Gaia archive website is \href{https://archives.esac.esa.int/gaia}{https://archives.esac.esa.int/gaia}.

%%%%%%%%%%%%%%%%%%%%%%%%%%%%%%%%%%%%%%%%%%%%%%%%%%%%%%%%%
%
%
%
%                Section: Appendix
%
%
%
%%%%%%%%%%%%%%%%%%%%%%%%%%%%%%%%%%%%%%%%%%%%%%%%%%%%%%%%%

\appendix

\section{Combining the Model Aspects into a Global PDF \label{sec:appendix_model}}

\subsection{Global Milky Way Disk model}
We show here how the different model aspects presented in subsections (\ref{subsection:pLzo}, \ref{subsection:p_Lz}, \ref{section_enrichment}, \ref{subsection:p_heating}, \ref{subsection:p_z}) are combined together to build the overall model for the Milky Way disk. Applying the probabilistic chain rule, and marginalizing over the dummy variable \Lzo, the different aspects of the model appear:
\begin{equation}
\label{eq_p_model}
\begin{split}
p_
\mathrm{MW}(\tau, \fe, J_R, \Lz, z |\ppm) &= 
 \int p(\Lzo | \ppm) p(\tau|\Lzo, \ppm)
 p(\fe, J_R, \Lz, z |\Ro, \tau, \ppm) \mathrm{d}\Lzo\\
& = \int p(\Lzo|\ppm)p(\tau|\Lzo,\ppm)
 p(\fe|\Lzo, \tau,\ppm)p(z |\Lzo, \tau,\ppm)
 p(J_R, \Lz |\Lzo, \tau,\ppm) \mathrm{d}\Lzo.
\end{split}
\end{equation}
The first term on the right hand side is the stars' birth angular momentum distribution (the first part of Subsection~\ref{subsection:pLzo}). The second is the star formation history conditioned on birth angular momentum, resulting from an inside-out star formation history (second part of Subsection~\ref{subsection:pLzo}). The third term is the distribution of metallicity in the star forming disk in function of time (modeled here as a Dirac function, since we are assuming a tight $\Lzo-\tau-\fe$ relation, Subsection~\ref{section_enrichment}), the fourth term is the vertical distribution of stars in the disk, and the last one is the joint distribution of in-plane orbital properties (which we take as the azimuthal action, or angular momentum, and the radial action). It can be split
\begin{equation}
\begin{split}
p(J_R,& \Lz~|~\Lzo, \tau,\ppm) =  p(J_R ~|~ \Lz, \Lzo, \tau,\ppm)p(\Lz~|~\Lzo, \tau,\ppm),
\end{split}
\end{equation}
where the first part corresponds to radial heating, and is conditioned on both birth angular momentum and present-day angular momentum. Heating through scattering should happen over the entire trajectory of the star, so in some sense at an average of the birth and final angular momenta, but we will drop the dependence on birth angular momentum as an approximation (see Subsection~\ref{subsection:p_heating}). The second term corresponds to radial migration, modelled as diffusion in angular momentum. Here it shows the probability of a star to be at angular momentum $L_z$ given it was born at $\Lzo$ a time $\tau$ ago (see Subsection~\ref{subsection:p_Lz}).
Each of these model aspects are presented in Section~\ref{sec:model} and assembled together to form the Milky Way model in Section~\ref{sec:likelihood}.

\subsection{Modeling the Dataset: Noise Model, Selection Function, and the Observables}
The Milky Way model described above cannot be directly applied to our dataset, since the stars were selected in a given survey and the data are noisy. We therefore write the model for the data set, in the space of the noisy observables (with subscript 'obs'). We (1) marginalize over uncertainties, and (2) apply the selection function to the model:
\begin{equation}
\begin{split}
    p&_\mathrm{dataset} = p_\mathrm{dataset}(l, b, D_\mathrm{obs}, v_{X,\mathrm{obs}}, v_{Y,\mathrm{obs}}, \fe_\mathrm{obs},
    \tau_\mathrm{obs} | \ppm, \mathbf{\sigma}) \\
    & = \int p_\mathrm{dataset}(l, b, D_\mathrm{true}, v_{X,\mathrm{true}}, v_{Y,\mathrm{true}}, \fe_\mathrm{true},
    \tau_\mathrm{true}|\ppm) p_\mathrm{noise}(\mathrm{obs}~|~\mathrm{true}, \mathbf{\sigma}) \mathrm{d}^n\mathrm{true}\\
    & = \frac{1}{V_s(\ppm)} \int p_\mathrm{MW}(l, b, D_\mathrm{true}, v_{X,\mathrm{true}}, v_{Y,\mathrm{true}}, \fe_\mathrm{true},
    \tau_\mathrm{true}|\ppm) S(l, b, D_\mathrm{true})f_{RC}(\tau_\mathrm{true}) p_\mathrm{noise}(\mathrm{obs}~|~\mathrm{true}, \mathbf{\sigma}) \mathrm{d}^n\mathrm{true}\\
    & = \frac{1}{V_s(\ppm)}\frac{1}{(2\pi)^2}\int p_\mathrm{MW}(L_z, J_R, z,\fe_\mathrm{true},
    \tau_\mathrm{true}|\ppm) D_\mathrm{true}^2 \cos(b)S(l, b, D_\mathrm{true})f_{RC}( \tau_\mathrm{true})p_\mathrm{noise}(\mathrm{obs}~|~\mathrm{true}, \mathbf{\sigma}) \mathrm{d}^n\mathrm{true}.
\end{split}
\end{equation}
From the first to the second line, we marginalize over data uncertainties with a noise model $p_\mathrm{noise}(\mathrm{obs}~|~\mathrm{true}, \mathbf{\sigma})$ where `noise' denotes all the noisy variables used here (those which have subscript `obs'), and the uncertainty parameter array $\sigma$ reflects the uncertainties described in the Section~\ref{section_data}. From the second to the third line,  we split the dataset model as the product of the Milky Way model extensively described in Section~\ref{sec:model} and the selection function $S(l, b, D, \tau)$. From the third to the fourth line, we describe the disk in the 2D space of actions ($J_R, L_z$) instead of the 4D phase space ($x, y, v_x, v_y$). We note the slight inconsistency in the actual action calculation based on the Staeckel approximation \citep{binney_2012} which uses the full 6D phase space information, and our simplified modelling assumption based on the adiabatic distribution functions \citep{binney_2010}. This assumption should not matter much since we restrict to the thin ($|b|<25$ deg), young (red clump), low-$\alpha$ disk where vertical excursions are very limited. The $D_\mathrm{true}^2 \cos(b)$ term is the Jacobian to change from Galactic to Cartesian coordinates.
In practice, we perform this integral by sampling the true values (in phase space and propagating directly to action space) from an approximate noise model $p(\mathrm{true}|\mathrm{obs})$, and then we use importance sampling (weighting the integrand with $p(\mathrm{obs}|\mathrm{true})/p(\mathrm{true}|\mathrm{obs})$) and Monte Carlo integrate by summing the remaining terms over these samples. The $p(\mathrm{obs}|\mathrm{true})$ term is the uncertainty model as described in Section~\ref{section_data}.

\section{Accounting for the Survey Volume $V_S(\ppm)$ \label{sec:appendix_Vs}}

The survey volume in Eq.~\ref{eq:survey_volume} is a 7 dimensional integral (or 5, after having integrated over $(l, b)$ by assuming the distribution function does not vary over an individual APOGEE field). We choose to compute this integral by importance sampling. This method works best if the proposal distribution $p_\mathrm{proposal}$ is similar to the target distribution. Ideally, we would like to generate 
from our full models
using a set of realistic parameters $\mathbf{p_\mathrm{prop}}$. We could then systematically use these samples in a Monte Carlo integration of the survey volume corresponding to a new set of model parameters. 
The advantage of this method is that our samples approximately trace the best fit model so provide an accurate computation of the normalization with a minimal number of samples. It also provides a tractable way to handle the selection function in the survey volume which is automatically incorporated in our sampling distribution. 

However, one cannot sample points directly from our full models as because they are constructed in a complex and un-normalized way. As a workaround, we choose to sample from the full model in two steps, using a simpler Galaxy model as an intermediate distribution. 
First, we sample stars from the simple proposal distribution that is easy to normalize. Then, we down-sample these data through importance sampling using our proposal model.
The simple Galaxy model $p_s$ is an exponential disk of constant scale length $R_{ds}$ and scale-height $h_{zs}$ chosen close to the analogous parameters in the overall model (e.g. $R_{d,\mathrm{prop}}$). Given their positions in the disk, stars velocities are sampled from a Gaussian centered on $(v_R,v_\phi)=(0,v_\mathrm{circ}(R))$ with large standard deviations that envelope the known velocity dispersion. 

To generate $N_\mathrm{prop}$ samples from our proposal distribution, we use the following procedure:
\begin{enumerate}
    \item For every APOGEE field $\mathrm{i}$, we sample on-sky positions $(l, b)$ using boundaries defined by the selection function. We then sample distances $D$ using the cumulative distribution function of
    \begin{equation}\label{eq:appendix_simple_proposal}
    p_s(D~|~l, b) \sim D^2 \cos(b) \exp(-R(l, b, D)/R_{ds}) \mathrm{sech}^2(z(l,b,D)/h_{zs}).
    \end{equation}
   
    \item We down-sample from these positions using the relative normalization of the simple model in each field.
    The field that contains the greatest number of stars is not down-sampled, and the other fields are down-sampled by accepting the points with probability
    \begin{equation}
    P_\mathrm{accept}(\mathrm{point}~|~ \mathrm{field~i}) =F_\mathrm{proposal}(l_\mathrm{i}, b_\mathrm{i})/F_\mathrm{max},
    \end{equation}
    where
    \begin{equation}
    F_\mathrm{proposal}(l_\mathrm{i}, b_\mathrm{i}) = \int_{D_{\mathrm{min,i}}}^{D_{\mathrm{max,i}}} p_s(D~|~l_\mathrm{i}, b_\mathrm{i})S(l_\mathrm{i}, b_\mathrm{i}, D) \mathrm{d}D
    \end{equation}
    and $F_\mathrm{max}=\mathrm{max}(F_\mathrm{proposal}(l_\mathrm{i}, b_\mathrm{i}))$. After this step, the samples have a distribution that follows our simple Galaxy model but now accounts fully for the spatial selection function of APOGEE and the 3D extinction.
    
    \item We now sample the remaining variables from our simple model before a further downsampling. Ages are sampled from a uniform distribution. Birth angular momenta \Lzo~are sampled from a normal distribution centered on $R\times 235$ kpc km/s with a standard deviation that increases as $\sigma \sim \sqrt{\tau}$ (to mimic radial migration). The strength $\sigma$ is chosen larger than the analogous parameter in the global model (\srm) such that this distribution envelopes the radial migration model. Radial and azimuthal velocities ($v_R$, $v_\phi$) are sampled from normal distributions centered on $0$ and $v_\mathrm{circ}(R)$ respectively. 
    % This makes a model that is normalize-able and normalized $p(l, b, D, \Lzo, v_r, v_\phi)$ inside APOGEE fields, with a Galactic component $p_s(\Lzo, R, z, v_r, v_\phi)$.
    \item We further down-sample these points by accepting them with a probability 
    \begin{equation}
    P_\mathrm{accept} =\frac{p_\mathrm{proposal}(L_z, J_R, \tau, \Lzo, z~|~\mathbf{p_\mathrm{prop}})f_{RC}(\tau)}{p_s(\Lzo, R, z, v_r, v_\phi)},
    \end{equation}
    where $p_\mathrm{proposal}(L_z, J_R, \tau, \Lzo, z~|~\mathbf{p_\mathrm{prop}})$ is our full proposal model with fixed parameters $\mathbf{p_\mathrm{prop}}$. The term $f_{RC}(\tau)$ contains the age-dependent fraction of stars on the red clump evolutionary stage.
    This down-sampling leads to a set of $N_\mathrm{prop}$ points.
\end{enumerate}
 
Our $N_\mathrm{prop}$ samples are now drawn from the pdf $S(l,b,D)f_{RC}(\tau) p_\mathrm{proposal}(L_z, J_R, \tau, \Lzo, z~|~\mathbf{p_\mathrm{prop}})$, so we can calculate the survey volume as
\begin{equation} \label{eq:survey_volume_appendix}
    \begin{split}
    V_S(\ppm) &= \int _\mathcal{D} p_\mathrm{model}(\mathcal{D} ~|~\ppm) S(l,b,D)f_{RC}(\tau) \mathrm{d}\mathcal{D}    \\    
    & = \int_\mathcal{D} \frac{p_\mathrm{model}(L_z, J_R, \tau, \Lzo, z ~|~\ppm) }{p_\mathrm{proposal}(L_z, J_R, \tau, \Lzo, z~|~\mathbf{p_\mathrm{prop}})} \times S(l,b,D) f_{RC}(\tau) p_\mathrm{proposal}(L_z, J_R, \tau, \Lzo, z~|~\mathbf{p_\mathrm{prop}})  \mathrm{d}\mathcal{D}\\
    & \approx \frac{1}{N_\mathrm{prop}}\sum_i^{N_\mathrm{prop}} \frac{p_\mathrm{model}(L_{z\mathrm{prop}i}, J_{R\mathrm{prop}i}, \tau_{\mathrm{prop}i}, L_{z0\mathrm{prop}i}, z_{\mathrm{prop}i} ~|~\ppm) }{p_\mathrm{proposal}(L_{z\mathrm{prop}i}, J_{R\mathrm{prop}i}, \tau_{i\mathrm{prop}}, L_{z0\mathrm{prop}i}, z_{\mathrm{prop}i}~|~\mathbf{p_\mathrm{prop}})}. 
    \end{split}
    \end{equation}
This integration has several advantages over using a regular grid (which is inefficient as the number of dimensions in the data increases) or re-sampling the normalization sample each time with new parameters. First, we only need to produce Monte Carlo samples once, not each time we need to evaluate the model, which is computationally more efficient. Secondly, since the initially generated samples are fixed, we need not recompute actions from these samples each time the model is evaluated, which saves additional computation time. Thirdly, as highlighted by \cite{mcmillanbinney2013}, for fixed samples the stochastic noise is limited and the overall normalized model is a smooth function of the model parameters which is a desirable property when we want to optimize the likelihood to fit the model parameters).

Step 2 of our procedure (down-sampling from the points generated at step 1) is in principle not mandatory. But in practice, the more alike the two distributions in the integral are, the greater the effective sample size.

We have tested this integration method on mock data, and the results remain robust as long as the proposal distribution is broad enough to envelope the distribution that we want to normalize. Additionally, we have tested the overall optimization scheme (MCMC ran on a model using this integration method) on mock data, and recovered the true parameters largely within the uncertainties.

%%%%%%%%%%%%%%%%%%%%%%%%%%%%%%%%%%%%%%%%%%%%%%%%%%%%%%%%%
%
%
%
%                Section: Bibliography
%
%
%
%%%%%%%%%%%%%%%%%%%%%%%%%%%%%%%%%%%%%%%%%%%%%%%%%%%%%%%%%
\bibliography{lit}
\end{CJK*}
\end{document}